\documentclass[english,showpacs,aps,prc,floats,floatfix,nofootinbib,superscriptaddress,a4paper]{revtex4}
\usepackage[T1]{fontenc}
\usepackage[latin9]{inputenc}
\usepackage{amssymb}
\usepackage{graphicx}
\usepackage{color}
\usepackage{hyperref}

\makeatletter

\providecommand{\tabularnewline}{\\}
\@ifundefined{textcolor}{}
{%
 \definecolor{BLACK}{gray}{0}
 \definecolor{WHITE}{gray}{1}
 \definecolor{RED}{rgb}{1,0,0}
 \definecolor{GREEN}{rgb}{0,1,0}
 \definecolor{BLUE}{rgb}{0,0,1}
 \definecolor{CYAN}{cmyk}{1,0,0,0}
 \definecolor{MAGENTA}{cmyk}{0,1,0,0}
 \definecolor{YELLOW}{cmyk}{0,0,1,0}
}

\usepackage{bm}
\usepackage{graphics}
\usepackage{epsf}
\makeatother

\def\s#1{\setbox0=\hbox{$#1$}  
   \dimen0=\wd0     
   \setbox1=\hbox{/} \dimen1=\wd1  
   \ifdim\dimen0>\dimen1   
      \rlap{\hbox to \dimen0{\hfil/\hfil}} 
      #1     
   \else     
      \rlap{\hbox to \dimen1{\hfil$#1$\hfil}} 
      /      
   \fi}      %

\usepackage{babel}
\begin{document}

\title{The nucleon mass and pion-nucleon sigma term from a chiral analysis of lattice QCD world data}

\author{L. Alvarez-Ruso}
\email{alvarez@ific.uv.es}
\affiliation{Instituto de F\'isica Corpuscular (IFIC), Centro Mixto
Universidad de Valencia-CSIC, E-46071 Valencia, Spain}
\author{T. Ledwig}
\email{ledwig@ific.uv.es}
\affiliation{Departamento de F\'\i sica Te\'orica and IFIC, Centro Mixto
Universidad de Valencia-CSIC, E-46071 Valencia, Spain}
\author{J. Martin Camalich}
\email{j.camalich@sussex.ac.uk}
\affiliation{Department of Physics and Astronomy, University of Sussex,
BN1 9QH, Brighton, UK}
\author{M. J. Vicente-Vacas}
\email{vicente@ific.uv.es}
\affiliation{Departamento de F\'\i sica Te\'orica and IFIC, Centro Mixto
Universidad de Valencia-CSIC, E-46071 Valencia, Spain}

\begin{abstract}
The pion-mass dependence of the nucleon mass within the covariant $SU(2)$ baryon chiral
perturbation theory both without and with explicit $\Delta\left(1232\right)$ degrees of freedom up to
order $p^{4}$ is investigated. By fitting to  a comprehensive set of  lattice QCD data in 2 and 2+1 flavors from several collaborations, for
pion masses $M_\pi < 420$ MeV, we obtain low energy constants of natural size and
compatible with pion nucleon scattering data. Our results are consistent with the rather
linear pion-mass dependence showed by lattice QCD. In
the 2 flavor case we have also performed simultaneous fits to nucleon mass and
$\sigma_{\pi N}$ data. As a result of our analysis, which encompasses the study of 
finite volume corrections and discretization effects, we report a value of $\sigma_{\pi N}=41(5)(4)$~MeV  in the 2 flavor
case and $\sigma_{\pi N}=52(3)(8)$ MeV for 2+1 flavors, where the inclusion of the
$\Delta\left(1232\right)$ resonance changes the results by around $9$~MeV. In the 2
flavor case we are able to set independently the scale for lQCD data, given by a Sommer
scale of $r_0=0.493(23)$~fm.
\end{abstract}



\maketitle

\section{Introduction}
The nucleon mass $M_{N}$ is one of the fundamental observables in nature. It arises from the complex and not well understood quark-gluon dynamics in the non-perturbative regime of quantum chromodynamics (QCD).  Nevertheless, important progress arises from the interplay of Chiral Perturbation Theory ($\chi$PT), the effective theory of QCD at low energies~\cite{Weinberg:1978kz,Gasser:1983yg,Gasser:1987rb,Leutwyler:1993iq} and lattice QCD (lQCD)~\cite{Fodor:2012gf}, in spite of the technical difficulties to perform lQCD simulation for light-quark masses close to the physical values. This strategy allows to extract some of the parameters of $\chi$PT that may not be easily accessible in experiments, clarify the role of baryon resonances in the nucleon selfenergy and unravel its strangeness content~\cite{Jenkins:1990jv,Borasoy:1996bx}. 

A measure of the contribution from explicit chiral symmetry breaking to the nucleon mass is provided by the so-called sigma terms. In particular, the  pion-nucleon $\sigma_{\pi N}$-term is defined as 
\begin{equation}
  \sigma_{\pi N}=\overline{m}\langle N |\overline{u}u+\overline{d}d| N \rangle\,\,\,,\label{eq: Sigma Term}
\end{equation}
in the isospin limit $m_{u}=m_{d}=\overline{m}\approx 4$~MeV. Using the Hellmann-Feynman (HF) theorem, $\sigma_{\pi N}$ can be related to $M_N$~\cite{Hellmann,Feynman:1939zz,Gasser:1980sb} 
\begin{equation}
  \sigma_{\pi N}=\overline{m}\frac{\partial}{\partial\overline{m}}M_{N}\left(\overline{m}\right)\,\,\,\,.
\label{eq:HF}
\end{equation}
Additionally, $\sigma_{\pi N}$ is the nucleon scalar form factor coming from light quarks at zero four-momentum transfer squared. As such, it enters quadratically in the scattering cross section of supersymmetric dark-matter particles with nucleons. Uncertainties in the determination of sigma-terms, including $\sigma_{\pi N}$, currently represent the largest source of error in direct dark-matter searches~\cite{Bottino:2001dj,Bottino:2008mf,Ellis:2008hf}. 

Traditionally, the pion-nucleon sigma term has been isolated by extrapolating $\pi N$-scattering data to the (unphysical) Cheng-Dashen point ($t=2 M_{\pi}^{2}$, $s=u=M_N^2$, where $s$, $t$ and $u$ are the standard Mandelstam variables)~\cite{Cheng:1970mx} using dispersive techniques. The results over the past three decades, $\sigma_{\pi N} = 49 \pm 8$~\cite{Koch:1982pu}, $\simeq45$ ~\cite{Gasser:1990ce}, $56 \pm 9$~\cite{Olsson:1999jt}, $64 \pm 7$~\cite{Pavan:2001wz}, $66 \pm 6$~\cite{Hite:2005tg}, $43 \pm 12$~MeV~\cite{Stahov:2012ca} \,\, \footnote{In the case of Refs.~\cite{Koch:1982pu,Olsson:1999jt,Pavan:2001wz,Hite:2005tg}, from the published value of the sigma-term at the Cheng-Dashen point $\sigma_{\pi N}(t=2 M_{\pi}^{2})$ we have subtracted $\Delta_\sigma =\sigma_{\pi N}(t=2 M_{\pi}^{2}) - \sigma_{\pi N}= 15.2 \pm 0.4$ according to the dispersive analysis of Ref.~\cite{Gasser:1990ap}. Additionally, see Ref.~\cite{Dominguez:1981kq} for the $\sigma_{\pi N}$-status before 1981.} depend on the data used as input and on the extrapolation procedure. The lack of consistency among the data sets as well as discrepancies between the parametrizations of the experimental data are a sizable source of systematic uncertainties. 

In order to sort the systematic effects out, much effort has been made in the context of baryon $\chi$PT (B$\chi$PT). At a given order in the chiral expansion, B$\chi$PT allows to express both the nucleon mass (and $\sigma_{\pi N}$) and the $\pi N$-scattering amplitude in terms of the same unknown low-energy constants (LECs). The available experimental information on $\pi N$-scattering can be used to obtain these LECs. Such a program has encountered a number of difficulties. Unlike in the meson sector, in B$\chi$PT the power counting (PC) is violated by the presence of $M_N$ as a heavy scale that does not vanish in the chiral limit. As a consequence, the loop diagrams do not fulfill the naive chiral order dictated  by their topology~\cite{Gasser:1987rb}. The solution to this problem follows from noticing that the genuine non-analytic chiral corrections indeed verify the PC, while the breaking pieces are analytic and can be renormalized into the LECs. Different approaches have been developed, including non-relativistic heavy-baryon (HB)~\cite{Jenkins:1990jv}, and the fully covariant infrared (IR)~\cite{Becher:1999he} and extended-on-mass-shell (EOMS)~\cite{Gegelia:1999gf,Fuchs:2003qc} schemes. In HB$\chi$PT, it was found that the convergence problems in some kinematic regions render the fits insensitive to the leading-order contribution to  $\sigma_{\pi N}$. The poor convergence can be traced back to the fact that the HB limit modifies the analytic structure of the $\pi N$ amplitude~\cite{Buettiker:1999ap}. To overcome the problems of HB$\chi$PT, the covariant formulations were developed. In the IR approach, loop functions are split into an infrared singular part which fulfills the PC and a regular part, containing the PC-breaking terms and higher order ones, which is dropped. An important drawback is that the IR scheme introduces unphysical cuts~\cite{Becher:1999he} which can have disruptive effects in low-energy phenomenology~\cite{Holstein:2005db,Geng:2008mf}. After applying this method to $\mathcal{O}(p^4)$, Becher and Leutwyler concluded that the IR chiral representation of the $\pi N$-scattering amplitude is a good approximation only in the subthreshold region so that no reliable determination of the sigma term could be performed from data in the physical region~\cite{Becher:2001hv}. In the EOMS, the PC is restored by renormalizing the finite number of PC-breaking terms. In this way, the analytic structure of the theory is preserved. Two recent EOMS studies of $\pi N$ scattering at order $p^3$~\cite{Alarcon:2012kn} and $p^4$~\cite{Chen:2012nx} have achieved a good description of the data and improved convergence.

A different complication concerns the treatment of the $\Delta(1232)\,3/2^+$ resonance which is only $\sim 300$~MeV heavier than the nucleon and couples strongly to the $\pi N$ system. In B$\chi$PT, the $\Delta(1232)$ is often treated as a heavy state whose influence in the observables is encoded in some of the LECs but, aiming at a more realistic description, it has often been taken explicitly into account. In order to include the $\Delta(1232)$ as a degree of freedom one needs to define a suitable PC for the new scale $\Delta = M_\Delta -M_N$~\cite{Hemmert:1996xg,Hemmert:1997ye,Pascalutsa:2002pi}, and to treat the so-called consistency problem afflicting interacting spin-3/2 fields (see Refs.~\cite{Pascalutsa:1999zz,Pascalutsa:2000kd,Pascalutsa:2006up} and references therein). The importance of explicitly including the $\Delta(1232)$ in B$\chi$PT has been stressed by a recent analysis of the $\pi N$ scattering amplitude performed in the EOMS scheme ~\cite{Alarcon:2011zs,Alarcon:2012kn,Chen:2012nx}. It was shown that the inclusion of the $\Delta$ resonance in a covariant framework is essential for a reliable extrapolation to the  Cheng-Dashen point~\cite{Alarcon:2012kn} . The resulting values of $\sigma_{\pi N}$ are in the 40-60 MeV interval, depending on the partial-wave analysis used as input and in agreement with those obtained by dispersive techniques~\cite{Alarcon:2011zs}. Although a value of $\sigma_{\pi N}=59 \pm 7$ MeV~\cite{Alarcon:2011zs} becomes eventually favored on the grounds of consistency with $\pi N$ phenomenology, an important conclusion of these works is that further efforts are required to understand the possible systematic errors in the $\pi N$ scattering data. 
 
Another way towards the determination of the $\pi N$ sigma term is provided by lQCD studies. Two different procedures have been used. In the first one, the matrix element in Eq.~(\ref{eq: Sigma Term}) is directly obtained and extrapolated to the physical values of the quark masses. The second procedure consists of using Eq.~(\ref{eq:HF}), after a suitable extrapolation of lQCD results for $M_N$ down to the chiral limit. The latter has been favored because of the technical difficulties that arise in the direct determination of disconnected contributions to $\sigma_{\pi N}$. 

The last decade has witnessed an impressive development of lQCD simulations. Results with two fully dynamical light (as light as possible) degenerate fermions ($N_f=2$) or with two degenerated light and one heavy (close to the physical strange-quark mass) flavor ($N_f=2+1$) have become standard. Even a direct determination of $\sigma_{\pi N}$ for $N_f=2+1+1$ (including dynamical $c$-quarks) has been reported~\cite{Dinter:2012tt}. Baryon $\chi$PT provides a natural framework to  extrapolate lattice data for $M_{N}$ with heavy quarks down to the physical and chiral limits, provided that the quark masses are small enough to warrant its applicability. In the context of HB$\chi$PT with a cut-off regularization it was already realized that non-analytic terms were important~\cite{Leinweber:2000sa,Leinweber:2003dg,Bernard:2003rp}. The quark-mass dependence of $M_{N}$ has also been investigated with $SU(2)$ IR$\chi$PT to $\mathcal{O}(p^4)$ without explicit $\Delta$~\cite{Procura:2003ig} and using phenomenological information to constrain the input parameters. Baryon $\chi$PT also allows to take finite lattice-volume corrections into account, as it was done for $M_{N}$ in Ref.~\cite{AliKhan:2003cu}.  A more complete $\mathcal{O}(p^4)$ IR$\chi$PT study~\cite{Procura:2006bj} included the leading  $\mathcal{O}(p^3)$ contribution of the $\Delta$ resonance with the small-scale expansion and HB approximation. According to this work, the introduction of $\Delta(1232)$ as a propagating degree of freedom is not crucial for $M_{N}$. This is in contrast with the findings of Ref.~\cite{Pascalutsa:2005nd} made with the EOMS scheme up to $\mathcal{O}(p^3)$.

 More recently, the $M_\pi$ dependence of new $N_f=2$ lQCD data for $M_{N}$ has been investigated with HB$\chi$PT~\cite{Alexandrou:2008tn} and IR$\chi$PT without explicit $\Delta$ degrees of freedom~\cite{Ohki:2008ff,Bali:2012qs,Chowdhury:2012wa}. The results for $\sigma_{\pi N}$ range from 37 to 67~MeV. In the case of Ref.~\cite{Bali:2012qs}, a direct measurement of $\sigma_{\pi N}$~\cite{Bali:2011ks} was incorporated to the fit, which allowed to increase the precision. Furthermore, three new direct determinations of  $\sigma_{\pi N}$ have also been performed applying noise reduction techniques for a better determination of the disconnected contribution~\cite{Alexandrou:2012zz}.  

Several collaborations have pursued lQCD simulations of the masses and $\sigma_{\pi N}$ using $N_f = 2 + 1$ configurations~\cite{Bernard:2001av,Aubin:2004wf,Bernard:2007ux,Durr:2008zz,Aoki:2008sm,WalkerLoud:2008pj,Lin:2008pr,Durr:2011mp,Bietenholz:2011qq,Beane:2011pc,Horsley:2011wr,Jung:2012rz}. The extrapolation to the physical point allows to determine $\sigma_{\pi N}$ together with other sigma terms and strangeness content of baryons. The difficulties encountered in HB$\chi$PT~\cite{WalkerLoud:2008pj,Ishikawa:2009vc} to accomplish this program were overcome applying cut-off regularization schemes~\cite{Young:2009zb,Shanahan:2012wh}, using covariant formalisms up to $\mathcal{O}(p^3)$~\cite{Semke:2006hd,MartinCamalich:2010fp,Geng:2011wq} and $\mathcal{O}(p^4)$~\cite{Semke:2011ez,Semke:2012gs,Ren:2012aj,Lutz:2012mq,Ren:2013dzt}, or complementing HB$\chi$PT with an expansion in the inverse number of colors (large-$N_C$)~\cite{Jenkins:2009wv,WalkerLoud:2011ab,Lutz:2012mq}. Although $SU(3)$-flavor calculations have reached a considerable degree of maturity, the large number of unknown LECs at $\mathcal{O}(p^4)$ and the size of the current lQCD data set limits, at present, on the accuracy attainable in the sigma terms. 

Alternatively, $SU(2)$ B$\chi$PT can be used to perform extrapolations of $M_N$ and $\sigma_{\pi N}$ in the light-quark masses with the implicit assumption that the influence of the strange quark is embedded in the LECs and that its mass in the simulations is close enough to the physical one. The chiral expansion is expected to converge faster than in $SU(3)$ B$\chi$PT and the different LECs appearing at $\mathcal{O}(p^4)$ can be independently determined using $\pi N$ scattering. On the other hand, in comparison with the $N_f=2$ simulations in which the strange quark is quenched, the extrapolated quantities from $N_f=2+1$ should be closer to those in the physical world. Analyses of $N_f=2+1$ simulations with $SU(2)$-HB$\chi$PT ansatzes at $\mathcal{O}(p^3)$ and without $\Delta(1232)$ have become standard~\cite{WalkerLoud:2008bp,Ishikawa:2009vc,Alexandrou:2009qu}. In particular, with HB$\chi$PT up to $\mathcal{O}(p^4)$ it was found that $\sigma_{\pi N} = 84 \pm 17 \pm 20$~MeV with explicit inclusion of the $\Delta$ resonance, and $\sigma_{\pi N} = 42 \pm 14 \pm 9$~MeV without it~\cite{WalkerLoud:2008bp}. While the inclusion of the $\Delta$ had little impact on the value of nucleon mass in the chiral limit, the central value of the sigma term changed by a factor of 2. It was also pointed out that the lattice data exhibited a surprisingly linear dependence on $M_\pi$, a feature also shown by other lQCD data~\cite{WalkerLoud:2008pj}. The importance of the $\Delta(1232)$ in extrapolations of lQCD data on $M_N$ has also been recently stressed in an analysis combining B$\chi$PT and the large-$N_c$ expansion.~\cite{CalleCordon:2012xz}. Finally, a different strategy was adopted in Ref.~\cite{Chen:2012nx}, according to which the LECs in $SU(2)$ B$\chi$PT up to $\mathcal{O}(p^4)$ were determined in simultaneous fits to $\pi N$ scattering data and lQCD results. 

Here we present our study of the pion mass dependence of the nucleon mass in covariant $SU(2)$ B$\chi$PT up to $\mathcal{O}(p^4)$, using the EOMS scheme with explicit inclusion of $\Delta(1232)$ intermediate states. We perform global fits to recent determinations of $M_N$ in lQCD simulations with $N_f=2$ and $N_f = 2 + 1$ dynamical quarks, taking into account finite lattice-volume corrections. By extrapolating the fits we determine the nucleon mass in the chiral limit and the pion-nucleon sigma term, paying attention to the different sources of systematic errors: the extrapolation to the continuum of lQCD data with finite lattice spacing, normalization errors, the uncertainties in the LECs fixed in the fits and the range of applicability of the chiral expansion. The article is organized as follows. In Sec.~\ref{sec:formalism} we describe the formalism, derive the formula for the nucleon mass and discuss the origin of the different coupling constants and LECs that are constrained in the fits. Finite volume corrections and continuum extrapolations are also discussed. The fit strategies and the results are presented in Sec.~\ref{sec:results}. We conclude and summarize our work in Sec.~\ref{sec:conclusions}. Further details about the calculation can be found in the Appendices.     

\section{Nucleon mass in the \textmd{B$\chi$PT}}
\label{sec:formalism}

Our aim is to study the pion mass ($M_{\pi}$) dependence of the nucleon
mass ($M_{N}$) and obtain the
value of the $\sigma_{\pi N}$-term by means of the HF theorem. 
For this we employ the $\mathcal{O}\left(p^{4}\right)$
covariant $SU\left(2\right)$ B$\chi$PT with and without explicit
$\Delta$-isobar degrees of freedom, $\Delta$B$\chi$PT and $\s\Delta$B$\chi$PT.
The resulting function $M_{N}\left(M_{\pi}\right)$ depends on several
LECs whose values we fix by fitting lQCD nucleon
mass data for unphysical quark-masses. The required ingredients are established in this Section.
We derive the perturbative nucleon mass and show the explicit fit formulas together with a discussion of lQCD
discretization effects. 

\begin{figure}
  \begin{centering}
    \includegraphics[scale=0.4]{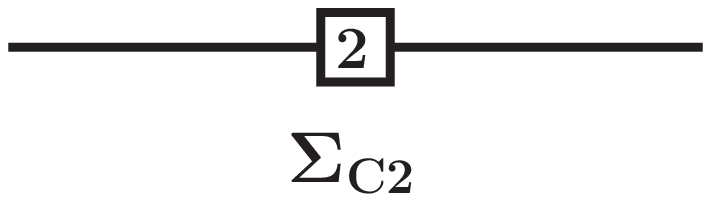}~~\includegraphics[scale=0.4]{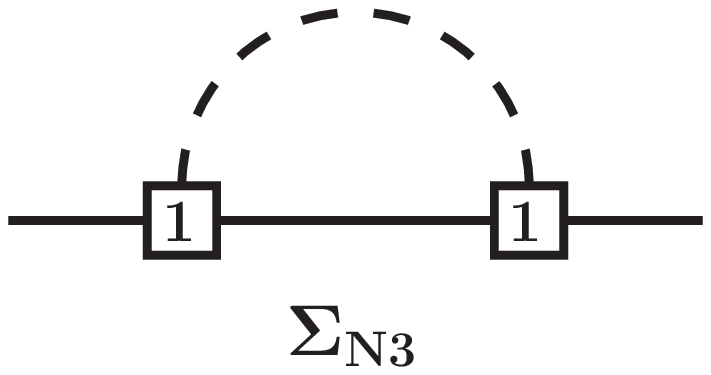}~~\includegraphics[scale=0.4]{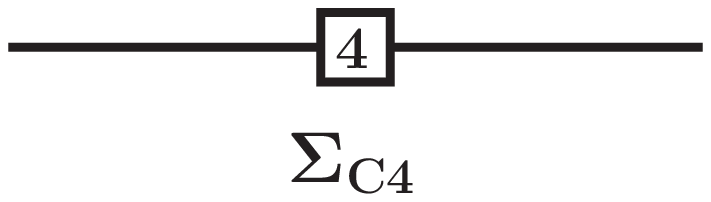}~~\includegraphics[scale=0.4]{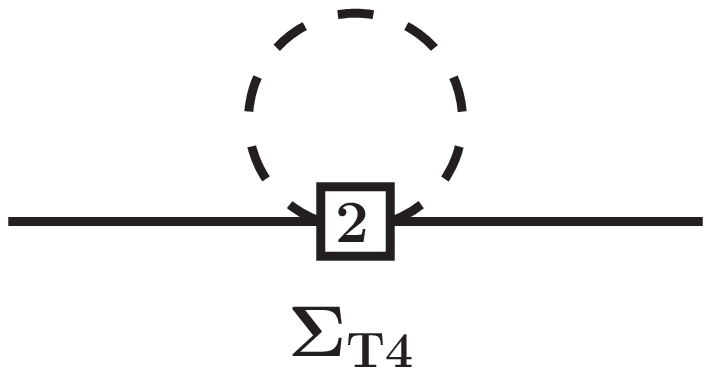}~~\includegraphics[scale=0.4]{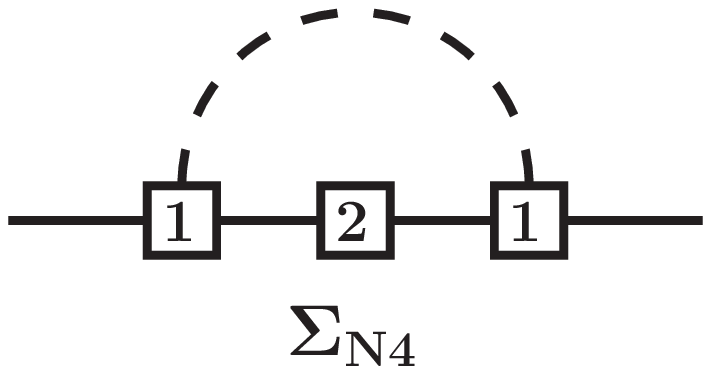}\\
    \includegraphics[scale=0.4]{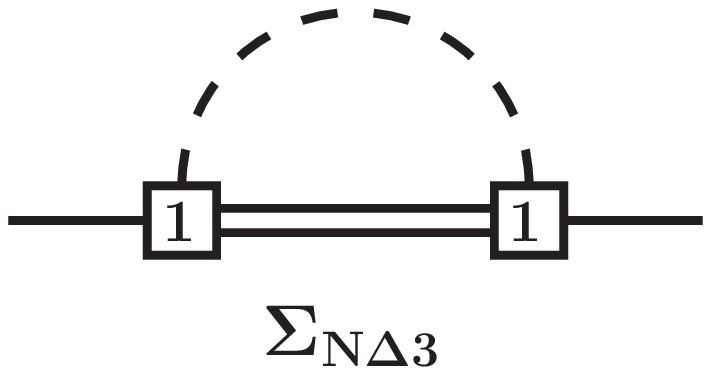}~~\includegraphics[scale=0.4]{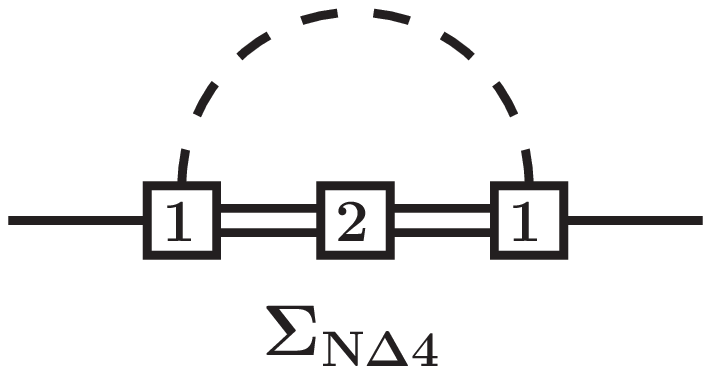}~~\includegraphics[scale=0.4]{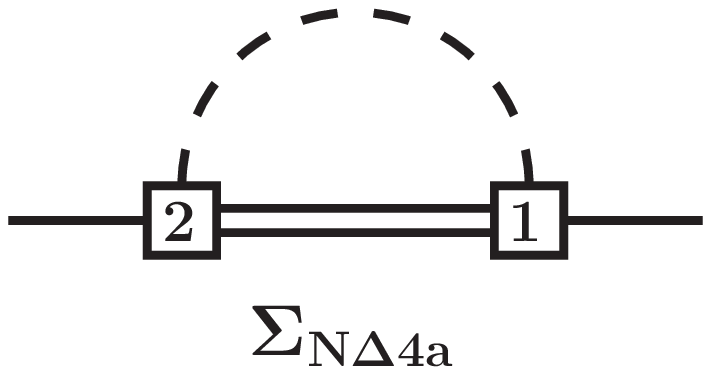}~~\includegraphics[scale=0.4]{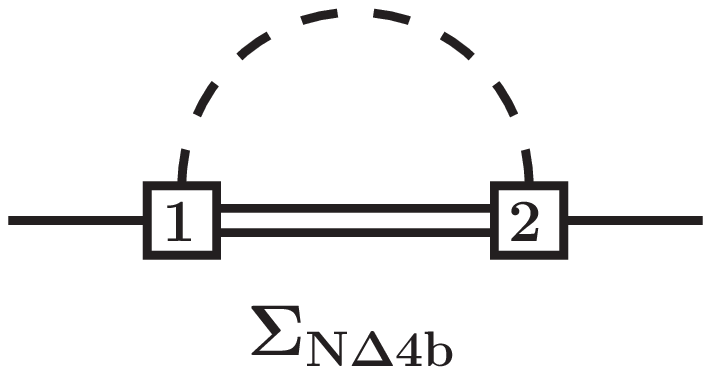}
    \par\end{centering}

    \caption{\label{fig: p4 nucleon diagrams}One-particle irreducible contributions to the
      nucleon self-energy up to $\mathcal{O}\left(p^{4}\right)$. Single solid lines denote nucleons, double
      solid lines, $\Delta$-isobars and dashed lines, pions. Boxes represent the pion-nucleon and contact vertices where the number
      specifies the chiral order.}
\end{figure}

To define the nucleon mass in terms of an expansion in the light scales $m_{\pi}^2 \equiv 2 B \overline{m}$, $p$ and $\Delta \equiv M_\Delta - M_N$, we have to choose a counting scheme\footnote{The constant $B$ is proportional to the chiral quark condensate.}.
If $\Delta$-isobars appear explicitly, the common assumptions 
are the small-scale expansion~\cite{Hemmert:1996xg,Hemmert:1997ye} that counts $\Delta \sim p \sim m_\pi$ and the $\delta-counting$~\cite{Pascalutsa:2002pi}, which takes $\Delta \sim p^{1/2}$ to preserve the hierarchy $p \sim m_\pi \ll \Delta$. As the latter is not the case for most of the lQCD simulations, we adopt the small-scale counting. The order $n$ of a self-energy contribution is then defined by 
\begin{equation}
  n=4L-2N_{\pi}-N_{N}-N_{\Delta}+\sum_{k}kV_{k}\,\,\,\,,\label{eq: Power Counting}
\end{equation}
for a graph with $L$ loops, $N_{\pi}$ internal pions, $N_{N}$ internal
nucleons, $N_{\Delta}$ internal $\Delta$-isobars and $V_{k}$ vertices
from a $\mathcal{L}^{\left(k\right)}$ Lagrangian. In Fig. \ref{fig: p4 nucleon diagrams}
we collect all one-particle irreducible diagrams that fulfill, after a suitable
renormalization, Eq. (\ref{eq: Power Counting}) up to $n =4$ [$\mathcal{O}(p^4)$] 
and list in App.~\ref{Appendix: BChPT-Lagrangians} all relevant B$\chi$PT Lagrangians.
Among the $\Delta$-isobar contributions, the graphs $\Sigma_{N\Delta4a}$ and $\Sigma_{N\Delta4b}$ originate from the $\mathcal{L}_{\pi N\Delta}^{\left(2\right)}$ Lagrangian \cite{Alarcon:2012kn}. It was shown in Ref.~\cite{Long:2010kt} 
for the HB$\chi$PT case that these couplings are redundant and can
be absorbed in the LECs of $\mathcal{L}_{\pi N}^{\left(2\right)}$
and $\mathcal{L}_{\pi N\Delta}^{\left(1\right)}$. The HB$\chi$PT
expressions are the leading order contributions to covariant B$\chi$PT
results which implies that these two diagrams start to contribute at $\mathcal{O}\left(p^{5}\right)$. 
We do not include them in our $\mathcal{O}\left(p^{4}\right)$
calculation. Additionally, the $\pi N$ scattering analysis \cite{Alarcon:2012kn}
performed explicitly fits with and without these terms and found strong
arguments to support that these redundancies also carry over to the covariant
case.

To calculate the remaining diagrams we apply the EOMS renormalization-scheme
\cite{Gegelia:1999gf,Fuchs:2003qc} which uses the analyticity of the power-counting
breaking terms to overcome the power-counting problem found in \cite{Gasser:1987rb}.
Explicitly, we calculate these diagrams in the dimensional regularization
for $D=4-2\epsilon$ dimensions and renormalize terms proportional
to  $L=-\frac{1}{\varepsilon}+\gamma_{E}-\ln4\pi$ ($\overline{MS}$-scheme) .
Subsequently, we renormalize the appearing LECs in such a way that power-counting breaking terms are canceled. 

\subsection{Nucleon self-energy and the perturbative nucleon mass }

The nucleon physical mass $M_{N}$ is defined by the pole position at $\s p=M_{N}$ of its full propagator 
\begin{equation}
  \frac{1}{\s p-M_{0}-\Sigma\left(\s p\right)}\,\,\,\,,
\end{equation}
where $\Sigma\left(\s p\right)$ and $M_{0}$ are the nucleon self-energy
and the (chiral limit) bare mass. In order to define a perturbative nucleon
mass, we expand $\Sigma\left(\s p\right)$ around $\s p=M_{0}$:
\begin{eqnarray}
  \Sigma\left(\s p\right) & = & \Sigma\left(M_{0}\right)+\left(\s p-M_{0}\right)\left.\frac{\partial}{\partial\s p}\right|_{\s p=M_{0}}\Sigma\left(\s p\right)+\frac{1}{2}\left(\s p-M_{0}\right)^{2}\left.\frac{\partial^{2}}{\partial\s p^{2}}\right|_{\s p=M_{0}}\Sigma\left(\s p\right)+...\\
  & = & \Sigma\left(M_{0}\right)+\left(\s p-M_{0}\right)\Sigma^{\prime}\left(M_{0}\right)+R\left(\s p\right)\,\,\,\,,
\end{eqnarray}
and write the propagator as
\begin{eqnarray}
  \frac{1}{\s p-M_{0}-\Sigma\left(\s p\right)} & = & \frac{1}{\displaystyle \s p-M_{0}-\frac{\Sigma\left(M_{0}\right)}{1-\Sigma^{\prime}\left(M_{0}\right)}-\frac{R\left(\s p\right)}{1-\Sigma^{\prime}\left(M_{0}\right)}}\,\frac{1}{1-\Sigma^{\prime}\left(M_{0}\right)}\,\,\,\,.\label{eq:expanded propagator}
\end{eqnarray}
Equation (\ref{eq:expanded propagator}) defines now the nucleon mass by the pole at $\s p=M_{N}$ 
\begin{equation}
M_{N} =  M_{0}+ Z\, \Sigma\left(M_{0}\right) + Z\, R\left(M_{N}\right)\,\,\,,\label{eq: Nucleon mass before BChPT}
\end{equation}
together with its residue
\begin{equation}
Z  = \frac{1}{1-\Sigma^{\prime}\left(M_{0}\right)}\,\,\,.
\end{equation}
Using the B$\chi$PT self-energies up to order $p^{4}$ of App. \ref{Appendix: Self-energy-formulas}
gives:
\begin{eqnarray}
  \Sigma_{p^{4}}\left(\s p\right) & = & \Sigma^{\left(2\right)}+\Sigma^{\left(3\right)}\left(\s p\right)+\Sigma^{\left(4\right)}\left(\s p\right)\\
  & = & \Sigma^{\left(2\right)}+\Sigma^{\left(3\right)}\left(M_{0}\right)+\Sigma^{\left(4\right)}\left(M_{0}\right)+\left(\s p-M_{0}\right)\left[\Sigma^{\left(3\right)\prime}\left(M_{0}\right)+\Sigma^{\left(4\right)\prime}\left(M_{0}\right)\right]+R\left(\s p\right)\,\,\,,  \label{eq:Nucleon mass interm}\\
  Z & = & 1+\Sigma^{\left(3\right)\prime}\left(M_{0}\right)+\mathcal{O}\left(p^{3}\right)\,\,\,,
\end{eqnarray}
where the upper indices denote the chiral order. Only the contact term $\Sigma_{C2}=-4c_{1}m_{\pi}^{2}$ enters in $\Sigma^{\left(2\right)}$ so it does not depend on  $\s p$. Inserting Eq.~(\ref{eq:Nucleon mass interm}) in Eq. (\ref{eq: Nucleon mass before BChPT})
one gets the nucleon mass up to order $p^{4}$:
\begin{eqnarray}
  M_{N}^{\left(4\right)}\left(m_{\pi}^{2}\right) 
  & = & M_{0}+\Sigma_{C2}\left(m_{\pi}^{2}\right)+\Sigma_{N3}\left(m_{\pi}^{2}\right)+\Sigma_{N\Delta3}\left(m_{\pi}^{2}\right)\label{eq: Nucleon p4 mass}\\
  &  & +\Sigma_{N4}\left(m_{\pi}^{2}\right)+\Sigma_{T4}\left(m_{\pi}^{2}\right)+\Sigma_{C4}\left(m_{\pi}^{2}\right)+\Sigma_{C2}\left(m_{\pi}^{2}\right)\Sigma_{N3}^{\prime}\left(m_{\pi}^{2}\right)\nonumber \\
  &  & +\Sigma_{N\Delta4}\left(m_{\pi}^{2}\right)+\Sigma_{C2}\left(m_{\pi}^{2}\right)\Sigma_{N\Delta3}^{\prime}\left(m_{\pi}^{2}\right)+\mathcal{O}\left(p^{5}\right)\,\,\,,\nonumber 
\end{eqnarray}
where all loops are evaluated at $\s p=M_{0}$. The term $R\left(M_{N}\right)$
contributes only at $\mathcal{O}(p^5)$. The first line of Eq. (\ref{eq: Nucleon p4 mass}) corresponds to
the $p^{3}$ nucleon mass while the second and third lines are the
additional $p^{4}$ contributions; the notation of the different terms matches the one of the diagrams in Fig.~\ref{fig: p4 nucleon diagrams}. All $\Sigma_{i}$ are obtained from the Lagrangians in App. \ref{Appendix: BChPT-Lagrangians}
and are explicitly given in App. \ref{Appendix: Self-energy-formulas}.
There are 10 low energy constants, namely, $f_{\pi0}, g_{A0}, c_{1}, c_{2}, c_{3}, h_{A0}, M_{0}, M_{\Delta0}, c_{1\Delta}, \alpha$. Most of them are constrained by experimental data. More details about their treatment are given below.

\subsection{Nucleon mass, $\sigma_{\pi N}$-term and fit formula\label{sub:pion mass, fit formula}}

Applying the HF theorem
\begin{equation}
  \sigma_{\pi N}\left(m_{\pi}^{2}\right)=\overline{m}\frac{\partial}{\partial\overline{m}}M_{N}\left(\overline{m}\right)=m_{\pi}^{2}\frac{\partial}{\partial m_{\pi}^{2}}M_{N}\left(m_{\pi}^{2}\right)
\end{equation}
to Eq. (\ref{eq: Nucleon p4 mass}) one obtains, 
\begin{eqnarray}
  M_{N}^{\left(4\right)}\left(m_{\pi}^{2}\right) & = & M_{0}-c_{1}4m_{\pi}^{2}+\frac{1}{2}\alpha m_{\pi}^{4}+\Sigma_{loops}^{\left(3\right)+\left(4\right)}\left(m_{\pi}^{2},M_{0},M_{\Delta0},f_{\pi0},g_{A0},h_{A0},c_{i}\right)\,\,\,\,,\label{eq:MN(mpi) loop}\\
  \sigma_{\pi N}^{\left(4\right)}\left(m_{\pi}^{2}\right) & = & -4c_{1}m_{\pi}^{2}+\alpha m_{\pi}^{4}+m_{\pi}^{2}\frac{\partial}{\partial m_{\pi}^{2}}\Sigma_{loops}^{\left(3\right)+\left(4\right)}\left(m_{\pi}^{2},M_{0},M_{\Delta0},f_{\pi0},g_{A0},h_{A0},c_{i}\right)\,\,\,\,,\label{eq:sima_Npi(mpi)}
\end{eqnarray}
with $c_{i}=c_{1},c_{2},c_{3},c_{1\Delta}$. The $\sigma_{\pi N}^{\left(4\right)}$
can also be obtained from a direct calculation of the nucleon scalar form
factor Eq. (\ref{eq: Sigma Term}) at zero four-momentum transfer squared. 
We have checked that Eq. (\ref{eq:sima_Npi(mpi)}) can be mapped term by
term to such a calculation, i.e. that our formulas with full, non-expanded
loops fulfill the HF theorem. 

To apply Eqs. (\ref{eq:MN(mpi) loop},\ref{eq:sima_Npi(mpi)}) with
a $p^{4}$ accuracy, we cannot identify the physical (or lattice) pion mass $M_\pi$ with the lowest order $m_\pi$ ($M_{\pi}^{2}=m_{\pi}^{2}=2B\overline{m}$) but must take the next order into account. According to the well known expansion~\cite{Gasser:1983yg}:
\begin{equation}
  M_{\pi}^{2}\left(m_{\pi}^{2}\right)=m_{\pi}^{2}+\frac{2l_{3}^{r}\left(\Lambda^{2}\right)}{f_{\pi0}^{2}}m_{\pi}^{4}+\frac{1}{32\pi^{2}f_{\pi0}^{2}}m_{\pi}^{4}\ln\frac{m_{\pi}^{2}}{\Lambda^{2}}+\mathcal{O}\left(p^{6}\right)\,\,\,,\label{eq:Mpi(mpi)}
\end{equation}
where $l_{3}^{r}\left(\Lambda^{2}\right)$ is a renormalized scale-dependent
LEC coming from the meson $\chi$PT Lagrangian. Therefore 
\begin{eqnarray}
  M_{N}^{\left(4\right)}\left(M_{\pi}^{2}\right) & = & M_{0}-c_{1}4M_{\pi}^{2}+\frac{1}{2}\overline{\alpha}M_{\pi}^{4}+\frac{c_{1}}{8\pi^{2}f_{\pi}^{2}}M_{\pi}^{4}\ln\frac{M_{\pi}^{2}}{M_{0}^{2}}+\Sigma_{loops}^{\left(3\right)+\left(4\right)}\left(M_{\pi}^{2},M_{0},M_{\Delta0},f_{\pi},g_{A},h_{A},c_{i}\right)+\mathcal{O}\left(p^{5}\right)\,\,,\label{eq:MN(4)(Mpi)}\\
  \sigma_{\pi N}^{\left(4\right)}\left(M_{\pi}^{2}\right) & = & -4c_{1}M_{\pi}^{2}+\overline{\alpha}M_{\pi}^{4}-c_{1}\frac{8}{f_{\pi}^{2}}l_{3}^{r}\left(M_{0}^{2}\right)M_{\pi}^{4}+\frac{c_{1}}{8\pi^{2}f_{\pi}^{2}}M_{\pi}^{4}\ln\frac{M_{\pi}^{2}}{M_{0}^{2}}\nonumber \\
  &  & +M_{\pi}^{2}\frac{\partial}{\partial M_{\pi}^{2}}\Sigma_{loops}^{\left(3\right)+\left(4\right)}\left(M_{\pi}^{2},M_{0},M_{\Delta0},f_{\pi},g_{A},h_{A},c_{i}\right)+\mathcal{O}\left(p^{5}\right)\label{eq:sigma(Mpi)}\\
  \mbox{with} &  & \overline{\alpha}=\alpha+c_{1}\frac{16}{f_{\pi}^{2}}l_{3}^{r}\left(M_{0}^{2}\right)\,\,\,\,.
\end{eqnarray}
Equation (\ref{eq:MN(4)(Mpi)}) is our final formula for $\mathcal{O}\left(p^{4}\right)$
B$\chi$PT fits to lQCD data. The effect of Eq. (\ref{eq:Mpi(mpi)})
is an additional $\mathcal{O}\left(p^{4}\right)$ term proportional
to $c_{1}$ and a redefinition of $\alpha\to\overline{\alpha}$ which
will be a fit parameter. Furthermore, we adopt the physical values
of $f_{\pi}=92.4$ MeV and $g_{A}=1.267$ instead of the chiral limit
ones and set the renormalization scale to $\Lambda = M_{0}$. The differences
between the chiral limit and physical values are of order $p^{2}$ so that they start
to contribute at $\mathcal{O}\left(p^{5}\right)$. In the case of $\sigma_{\pi N}^{\left(4\right)}\left(M_{\pi}^{2}\right)$
we cannot absorb all terms proportional to $l_{3}^{r}\left(M_{0}^{2}\right)$ in the LECs
and shall need a numerical value for it. From the latest estimate of $\overline{l}_{3}\left(M_{\pi}\right)=\ln\Lambda_{3}^{2}/M_{\pi}^{2}$
at the physical point $\overline{l}_{3}\left(139\,\mbox{MeV}\right)=3.2\left(8\right)$~\cite{Gasser:1983yg,Colangelo:2010et} one has: 
\begin{eqnarray}
  l_{3}^{r}\left(\Lambda^{2}\right) & = & -\frac{1}{64\pi^{2}}\left(\overline{l}_{3}\left(M_{\pi}\right)+\ln\frac{M_{\pi}^{2}}{\Lambda^{2}}\right)=-\frac{1}{64\pi^{2}}\left(3.2\left(8\right)+\ln\frac{M_{\pi\left(phys\right)}^{2}}{\Lambda^{2}}\right)\,\,\,\,,
\end{eqnarray}
where we set $M_{\pi\left(phys\right)}=139$ MeV. 

\subsection{Low-energy constants, finite volume and lattice spacing effects}

After fixing $f_{\pi0}$ and $g_{A0}$, we discuss our treatment of the remaining eight LECs, $c_{1}$, $c_{2}$, $c_{3}$,
$M_{0}$, $\overline{\alpha}$, $M_{\Delta0}$, $h_{A0}$, and $c_{1\Delta}$. Generally, our fits depend very mildly on variations in $c_{2}$, $c_{3}$, $M_{\Delta0}$ and $h_{A}$. Furthermore, we observe that changes in $c_{1\Delta}$ 
are compensated by changes in $\overline{\alpha}$. Our strategy is, therefore, to 
fit $M_{0}$, $c_{1}$ and $\overline{\alpha}$ while keeping $c_{2}$, $c_{3}$, $M_{\Delta0}$, $h_{A}$ and
$c_{1\Delta}$ fixed. The nucleon-related LECs $c_{2}$ and $c_{3}$ are taken from the 
$\pi N$-scattering analysis of Ref.~\cite{Alarcon:2011zs}, performed with the same 
B$\chi$PT framework employed here. More specifically, we take as central values 
the average of the results of fits to the phase shifts from the Karlsruhe-Helsinki group (KA85) 
and the George Washington University group (WI08), accepting errors defined by their uncertainties and 
also by the result of the fit to  Matsinos phase shifts (EM06) (see Tables 1 and 2 of  Ref.~\cite{Alarcon:2012kn} )
\footnote{Further justification for this choice is given in the Results Section.}. The specific figures for 
both the $\s\Delta$-B$\chi$PT and $\Delta$-B$\chi$PT cases are given in Table~\ref{tab:LECsuncertainties}.

\begin{figure}
  \begin{center}
  \begin{minipage}[t]{0.45\columnwidth}
    \includegraphics[scale=0.40]{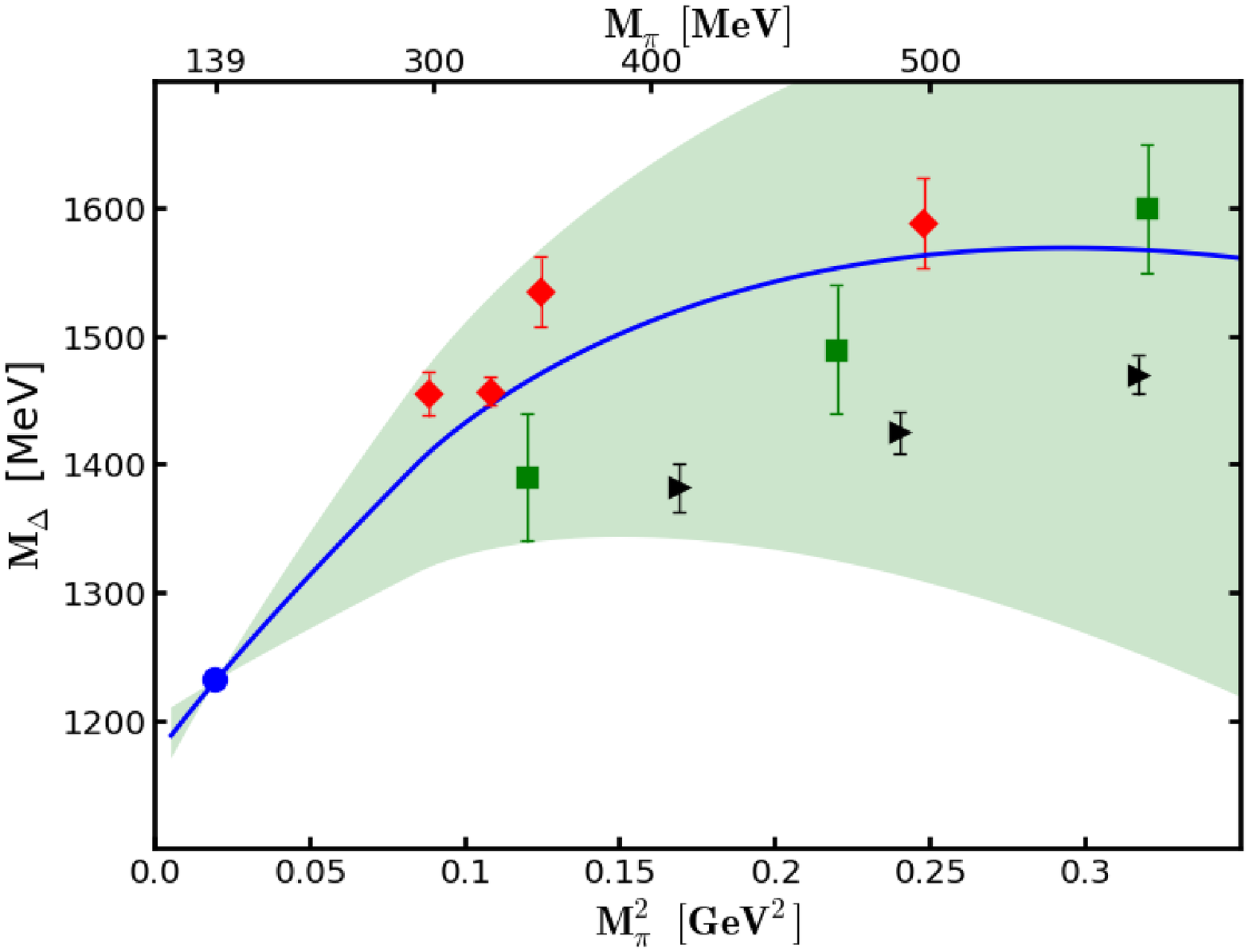}
    \caption{\label{fig:Pion-mass-dependence-Delta}Pion mass dependence of the $\Delta$-isobar mass. Green squares
      are from \cite{Bernard:2001av,Aubin:2004wf}, black right-triangles
      are quenched data from \cite{Alexandrou:2011py} and red
      diamonds are unquenched data from \cite{Alexandrou:2011py}.
      The blue circle is the physical point. The band defines the uncertainty range adopted (see the text) while the blue line is the preferred result.}
  \end{minipage}
  \hspace{0.05\columnwidth}
  \begin{minipage}[t]{0.45\columnwidth}
    \includegraphics[scale=0.40]{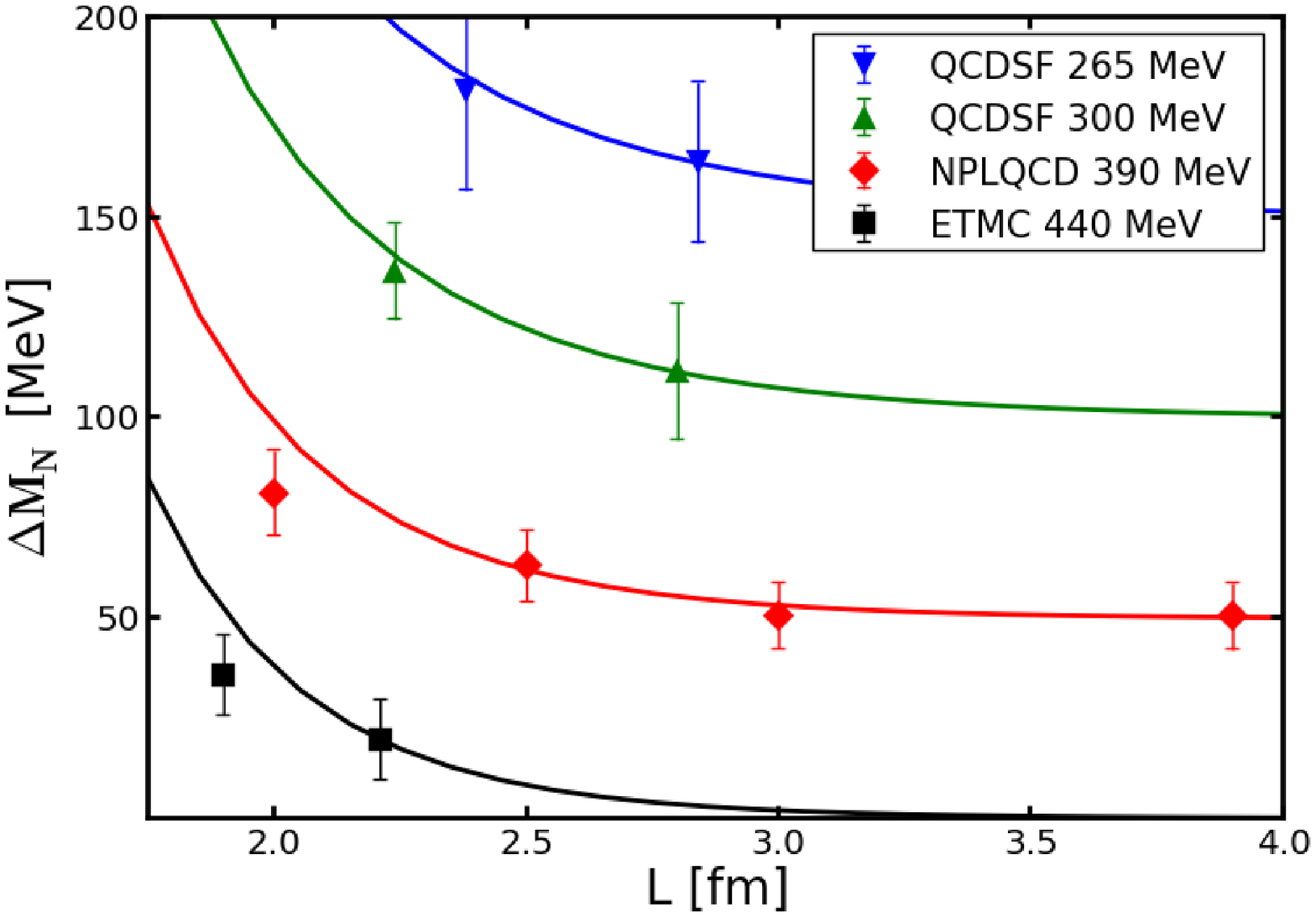}
    \caption{\label{fig:FV-test} Finite volume corrections $\Delta M_N = M_N(L) - M_N(L \rightarrow \infty)$ as a function 
of the lattice size for pion masses of $265$, $300$, $390$ and $440$ MeV. Lattice data from Refs.~\cite{Bali:2012qs} (triangles), \cite{Beane:2011pc} (red diamonds) and \cite{Alexandrou:2010hf} (squares) with approximately the same pion masses are also displayed.  We normalize each curve to the point with the
      largest volume and shifted them by multiples of $50$ MeV to avoid overlaps.
      At $L=4.0$ fm $\Delta M_N \approx 0$ for all curves.}
  \end{minipage}
  \end{center}
\end{figure}

In order to fix the $\Delta$-related LECs, $M_{\Delta0}$, $c_{1\Delta}$ and $h_{A}$, we consider the  
pion-mass dependence of the $\Delta$-isobar mass. Up to $\mathcal{O}\left(p^{3}\right)$ it reads~\cite{Pascalutsa:2005nd}
\begin{eqnarray}
M_{\Delta}^{\left(3\right)}\left(M_{\pi}\right) & = & M_{\Delta0}-4c_{1\Delta}M_{\pi}^{2}+\Sigma_{\Delta N3}\left(M_{\pi};\, h_{A},f_{\pi},M_{N},M_{\Delta}\right)+\Sigma_{\Delta\Delta3}\left(M_{\pi};\, H_{A},f_{\pi},M_{\Delta}\right)\,\,\,\,,\label{eq:MD mass with LECs}
\end{eqnarray}
where the loop contributions $\Sigma_{\Delta N3}$ and $\Sigma_{\Delta\Delta3}$ stand for diagrams like $\Sigma_{N3}$ and  $\Sigma_{N\Delta3}$ in Fig~\ref{fig: p4 nucleon diagrams} but with external nucleon lines replaced by $\Delta(1232)$ ones. The explicit expressions are given in  App. \ref{Appendix: Self-energy-formulas}.  As stated above, we are allowed to take phenomenological values for the LECs in these loops . In this way, one uses the phenomenological value of the $\Delta$-isobar decay width $\Gamma_{\Delta\to N\pi}=-2\,\mbox{Im}\Sigma_{\Delta N3}=115$~MeV to fix $h_{A}=2.87$. Furthermore, we adopt $H_{A}=\frac{9}{5}g_{A}$ obtained in the large-$N_c$ limit. Finally, we use lQCD data for the $\Delta(1232)$~\cite{Alexandrou:2011py,Bernard:2001av,Aubin:2004wf} mass to determine the remaining two LECs $M_{\Delta0}$ and $c_{1\Delta}$. As the available lattice results are rather scattered, we do not perform a rigorous fit to them but, instead, adopt the conservative attitude  of setting a band that englobes all the lQCD points with their errorbars (see Fig.~\ref{fig:Pion-mass-dependence-Delta}). The central values for the parameters result from the average of those defining the band's boundaries and are listed in Table~\ref{tab:LECsuncertainties}.

\begin{table}
  \begin{tabular}{|c||ccccc|}
    \hline 
    Theory & $c_{2}$~[GeV$^{-1}$]  & $c_{3}$~[GeV$^{-1}$] & $c_{1\Delta}$~[GeV$^{-1}$] & $h_{A}$ & $M_{\Delta0}$~[MeV]\tabularnewline
    \hline 
    \hline 
    $\s\Delta$-B$\chi$PT & $3.9\pm 0.4$ & $-6.7 \pm 0.4$ & $--$ & $--$ & $--$\tabularnewline
    $\Delta$-B$\chi$PT & $1.1^{+0.2}_{-0.5}$ & $-3.0^{+0.6}_{-0.1}$ & $-0.90 \pm 40$ & $2.87$ & $1170 \pm 30$\tabularnewline
    \hline 
  \end{tabular}
  \caption{\label{tab:LECsuncertainties} Values of the LECs appearing in the $p^{4}$
    nucleon mass. For the LECs $f_{\pi0}$
    and $g_{A0}$ we take their physical values $f_{\pi}=92.4$ MeV and
    $g_{A}=1.267$.}
\end{table}
 
We now turn  to two discretization artifacts: finite volume (FV) and finite spacing effects, appearing in lQCD studies, 
as a consequence of the finite grid with volume $L^{3}$ and spacing $a$ in which simulations are performed.

All loop graphs of Fig. \ref{fig: p4 nucleon diagrams} are subject to FV corrections. 
We calculate them  in App. \ref{Appendix: Finite-volume-expressions}
applying the standard techniques of Ref.~\cite{AliKhan:2003cu}. The FV corrections to $\Sigma_{N3}$
and $\Sigma_{T4}$ are equivalent to those in Ref.~\cite{AliKhan:2003cu}. In addition, we correct  
the combination $\Sigma_{N4}+\Sigma_{c2}\Sigma_{N3}^{\prime}$, the $\Delta$-isobar graphs $\Sigma_{N\Delta3}$, $\Sigma_{N\Delta4}$
and $\Sigma_{C2}\Sigma_{N\Delta3}^{\prime}$ which contribute at order $p^{4}$ in the continuum~\cite{Gegelia:1999gf,Fuchs:2003qc} 
in the EOMS renormalization scheme. Reference \cite{AliKhan:2003cu}
employs IR, for which the combination of $\Sigma_{N4}+\Sigma_{c2}\Sigma_{N3}^{\prime}$ appears only
at order $p^{5}$.  Our FV corrections are therefore:
\begin{eqnarray}
  \Sigma_{p^{4}}\left(M_{\pi}^{2},L\right) & = & \Sigma_{N3}\left(M_{\pi}^{2},L\right)+\Sigma_{N\Delta3}\left(M_{\pi}^{2},L\right)\nonumber \\
  &  & +\Sigma_{N4}\left(M_{\pi}^{2},L\right)+\Sigma_{N\Delta4}\left(M_{\pi}^{2},L\right)+\Sigma_{T4}\left(M_{\pi}^{2},L\right)\nonumber \\
  &  & +\Sigma_{C2}\left(M_{\pi}^{2}\right)\Sigma_{N3}^{\prime}\left(M_{\pi}^{2},L\right)+\Sigma_{C2}\left(M_{\pi}^{2}\right)\Sigma_{N\Delta3}^{\prime}\left(M_{\pi}^{2},L\right)\,\,\,.
\end{eqnarray}
All these terms are given in App. \ref{Appendix: Finite-volume-expressions}. In Fig. \ref{fig:FV-test} we test our FV
correction against lQCD data with approximately the same
pion mass but different $L$. We found four points from the QCDSF Collaboration \cite{Bali:2012qs},
four points from the NPLQCD Collaboration \cite{Beane:2011pc}
and two points from the ETM Collaboration \cite{Alexandrou:2010hf}
at pion masses approximately of $265$, $300$, $390$ and $440$
MeV, respectively.  Reasonable values of the LECs $M_{0}=890$~MeV and $c_{1}=c_{1\Delta}=-0.9$~GeV$^{-1}$ 
have been chosen for this exercise. We observe that our FV corrections describe very well the $L$ 
dependence for lattice sizes larger than $\sim2.2$ fm and that they have a size
of up to $45$ MeV. In our fits we shall include only data points with $LM_{\pi}>3.8$
for all of which $L > 2.2$ fm.

In general, we will use lQCD data that are not extrapolated to
the continuum limit $a\to0$. Originally, discretized QCD actions
break chiral symmetry even in the chiral limit by terms proportional
to $a$ \cite{Wilson:1974sk,Sheikholeslami:1985ij,Luscher:1996sc} but modern lattice calculations
use $\mathcal{O}\left(a\right)$ improved actions for which
discretization effects in baryon masses start at order $a^{2}$.
However, there exists a whole variety of lQCD-actions, each with its own  
discretization effects. For the specific Symanzik
lQCD action an effective field theory investigation has been performed in Ref.~\cite{Tiburzi:2005vy}
on a HB$\chi$PT basis but a general approach, similar to the treatment of FV corrections, 
does not exist. Therefore, we parametrize this effect for each action individually by writing the
nucleon mass in an $a$ -expansion to the lowest order as 
\begin{equation}
  M_{N}=M_{a=0}+c_{a}a^{2}+\mathcal{O}\left(a^{3},a^{2}m_{\pi}^{2}\right)\,\,\,\,,\label{eq:MN(a)}
\end{equation}
with an action-specific constant $c_{a}$. By using the ETMC points
at $M_\pi =260$ and 262~MeV, and QCDSF points at $r_{0}M_{\pi}=0.658$ and 0.660~\cite{Alexandrou:2010hf,Bali:2012qs} we can
roughly estimate the size of this effect. By taking the linear $a^{2}$-extrapolation
of Eq. (\ref{eq:MN(a)}) we obtain $c_{ETMC}=0.17$~GeV$^{3}$ and $c_{QCDSF}=0.33$ GeV$^{3}$, which correspond to nucleon-mass 
shifts of $10-50$ MeV. We obtain that lattice spacing corrections
can have similar sizes to the FV ones. Therefore, we incorporate 
this effect in specific fits by including the $c_{a}a^{2}$ term in the
$\chi^{2}$ for each collaboration/action reporting results for different values of $a$.

\section{Results}
\label{sec:results}

We study the pion mass dependence of the nucleon mass by using the
covariant B$\chi$PT expression of Eqs. (\ref{eq: Nucleon p4 mass}) and (\ref{eq:MN(4)(Mpi)}), 
which is accurate up to the chiral order $p^{4}$ and includes
explicit $\Delta$-isobar degrees of freedom. We perform global fits
to lQCD ensembles for $N_{f}=2$ and $N_{f}=2+1$ numbers of flavors.
Generally, lQCD uses a discretized QCD-action to simulate the quark-gluon
interaction in a finite box of size $L^{3}\times T$ with
finite spatial and time spacings of $a$ and $a_{t}$. The nucleon
mass data are given in terms of the dimensionless quantities $a M_{\pi}$
and $aM_{N}$ with uncertainties in $a$, $aM_{\pi}$ and $aM_{N}$.
An actual value of $a$ sets the overall scale to convert the lQCD data into physical units. 
No universal scale-setting method exists and different collaborations use different approaches.
Furthermore, the statistical uncertainty in $a$ turns into a normalization
uncertainty in $M_{N}$ for data points belonging to the same $a$-set.
It is therefore preferable to fit the $(aM_{\pi},aM_{N})$  data directly whenever this is possible or, otherwise, 
to include these correlated uncertainties in the fit. As explained below, we are able to perform the
former in the case of the $N_{f}=2$ ensembles and rely on the latter
for the $N_{f}=2+1$ ones. We also include FV corrections
and lattice spacing effects as described in the previous section.
We fit the LECs $M_{0}$, $c_{1}$ and $\overline{\alpha}$ while keeping 
$c_{2}$ , $c_{3}$, $c_{1\Delta}$, $h_{A}$ and $M_{\Delta0}$
fixed to the values listed in Tab. \ref{tab:LECsuncertainties}. Afterwards, we quantify the effect of varying
the fixed LECs within their ranges. The fit uncertainties are determined at a $68$\% confidence level.

For $N_{f}=2$ we include data from the BGR~\cite{Engel:2010my},ETMC~\cite{Alexandrou:2010hf}, Mainz~\cite{Capitani:2012gj} and QCDSF~\cite{Bali:2012qs} collaborations, and for $N_{f}=2+1$ from the BMW~\cite{Durr:2011mp}, HSC~\cite{Lin:2008pr}, 
LHPC~\cite{Bratt:2010jn}, MILC~\cite{MILC:2013}, NPLQCD~\cite{Beane:2011pc}, PACS~\cite{Aoki:2008sm} and RBCUK-QCD~\cite{Jung:2012rz} collaborations. In both cases we extract the LECs and obtain the $\sigma_{\pi N}$ value by using the HF theorem. 

\subsection{Nucleon mass up to order $\mathcal{O}\left(p^{4}\right)$: fits to
  $N_{f}=2$ lattice QCD data \label{sub:Nf2 fits}}

We use Eq. (\ref{eq:MN(4)(Mpi)}) to fit
the lQCD data for the $N_{f}=2$ ensembles of the BGR, ETMC, Mainz
and QCDSF collaborations \cite{Engel:2010my,Alexandrou:2010hf,Capitani:2012gj,Bali:2012qs}.
The lQCD data are given in terms of the dimensionless products $aM_{\pi}$
and $aM_{N}$ where the scale is fixed in different ways: with the experimental
$\Omega^{-}$ mass in Ref.~\cite{Capitani:2012gj} and with HB$\chi$PT or IR-$\chi$PT chiral extrapolations of $M_N$ in Refs.~\cite{Alexandrou:2010hf,Bali:2012qs}. The available information for these data sets is such that we can perform our own scale setting. By doing this we compensate for the different scales of the various sets and avoid manipulating them with two different B$\chi$PT versions.

Explicitly, we fit the lQCD data in terms of $\left(r_{0}M_{\pi},r_{0}M_{N}\right)$
by using the Sommer-scale $r_{0}$ \cite{Sommer:1993ce} and the ratios
$r_{0}/a$ in the chiral limit, as reported by each Collaboration.
The uncertainties in $aM_{\pi}$, $aM_{N}$ and $r_{0}/a$ are assumed to be uncorrelated. 
The value of $r_{0}$ is a priori unknown and we determine it recursively
inside the fit. This is the same strategy used in Ref.~\cite{Bali:2012qs}, now employed to analyze $N_{f}=2$ data globally. 
The $\chi^{2}$ function that we minimize is
\begin{eqnarray}
  \chi^{2} & = & \sum_{i}\left[\frac{\widetilde{M}_{N}^{(n)}\left(\widetilde{M}_{\pi}^{2}\right)+\widetilde{\Sigma}^{(n)}_{N}\left(\widetilde{M}_{\pi}^{2},L \right)+\tilde{c}_{a}\tilde{a}^{2}-d_{i}\left(\widetilde{M}_{\pi}^{2},L \right)}{\sigma_{i}}\right]^{2}\,\,\,\,,\label{eq:chi2 nf2}\\
  \mbox{with} &  & \widetilde{M}_{N}^{(n)}=r_{0}M_{N}^{(n)}\,\,\,\,,\,\,\,\,\widetilde{M}_{\pi}^{2}=\left(r_{0}M_{\pi}\right)^{2}\,\,\,\,,\,\,\,\,\widetilde{\Sigma}_{N}^{(n)}=r_{0}\Sigma_{N}^{(n)}\,\,\,\,,
\end{eqnarray}
where $d_{i}\left(\widetilde{M}_{\pi}^{2},L\right)$ are the 
lQCD data points with uncertainties $\sigma_{i}$, each of them generated in a lattice of size $L$ and spacing $a$.  
The continuum expressions $M_{N}^{(n)}\left(M_{\pi}^{2}\right)$
and the finite volume corrections $\Sigma^{(n)}\left(M_{\pi}^{2},L\right)$ for the chiral-order $n$
are listed in App. \ref{Appendix sub:Fit-formulas}. As discussed
above, the terms $\tilde{c}_{a}\tilde{a}^{2}=r_{0}^{3}c_{a}\left(a/r_{0}\right)^{2}$
parametrize discretization effects, with $c_{a}$ being common
constants for points obtained by the same lQCD Collaboration/action.
The Sommer-scale is calculated in each minimization step recursively 
using the constraint imposed by the experimental value of the nucleon mass at the physical
point: 
\begin{equation}
  r_{0}^{k}=\frac{\widetilde{M}_{N}^{\left(n\right)}\left(r_{0}^{k-1}\cdot M_{\pi\left(phys\right)}\right)}{M_{N\left(phys\right)}}\,\,\,\,\,\,\,\,\,\,\,\,\mbox{until}\,\,\,\,\,\,\,\,\,\,\,\,|r_{0}^{k}-r_{0}^{k-1}|<0.001\,\,\mbox{fm} \,\,\,\,.
\end{equation}
The explicit fit parameters in Eq. (\ref{eq:chi2 nf2}) are $M_{0}$, $c_{1}$, $\overline{\alpha}$ and two $c_{a}$ constants, one for the ETMC Collaboration and one for both Mainz and QCDSF which employ the same action. The single data point of BGR does not allow to perform any lattice spacing correction. As the term  $\tilde{c}_{a}\tilde{a}^{2}$ does not stand on the same firm ground, from the perspective of effective field theory, as the rest of our mass formula, we perform fit with and without it and treat the differences as systematic errors. We restrict the data sets by imposing the following conditions: $r_{0} M_{\pi} < 1.11$, $M_{\pi} L > 3.8$, which englobe points of $M_{\pi} < (429,476)$~MeV for Sommer-scale values in the range $r_{0}=(0.51,0.46)$~fm. We then consider the following data sets  
\begin{itemize}
\item BGR \cite{Engel:2010my}: A Sommer-scale of $r_{0}=0.48$ fm is
  assumed and three data points are provided, only one below $r_{0}M_{N}=1.11$. 
\item ETMC \cite{Alexandrou:2010hf}: Eleven data points are provided
  in the form $\left(aM_{\pi},aM_{N}\right)$; for each setting
  a value of $r_{0}/a$ is computed. After converting $\left(aM_{\pi},aM_{N}\right)$
  into $\left(r_{0}M_{\pi},r_{0}M_{N}\right)$ we find that seven data points fulfill our 
conditions and enter the fit. 
\item Mainz \cite{Capitani:2012gj}: Eleven data points are provided
  in the form $\left(aM_{\pi},aM_{N}\right)$. The lattice spacings
  as well as the ratios $r_{0}/a$ are determined by the $\Omega^{-}$
  mass~\cite{Capitani:2011fg,Leder:2011pz}. We
  convert $\left(aM_{\pi},aM_{N}\right)$ to $\left(r_{0}M_{\pi},r_{0}M_{N}\right)$
  and six data points enter the fit. 
\item QCDSF \cite{Bali:2012qs}: This work provides 27 data points, directly
  in terms of $\left(r_{0}M_{\pi},r_{0}M_{N}\right)$, but only two of them fulfill our 
restrictions. In addition, there is a single data point
for the $\sigma_{\pi N}$ obtained by direct determination at $M_\pi\sim285$ MeV~\cite{Bali:2011ks}. 
\end{itemize}

We study the following variations of the fits: 
\begin{enumerate}
\item $M_N (M_\pi)$ to order $p^2$, $p^3$ and $p^4$ in the chiral expansion
\item without ($\s\Delta$B$\chi$PT) and with ($\Delta$B$\chi$PT) $\Delta$-isobar
\item including and excluding the single direct $\sigma_{\pi N}$ measurement of Ref.~\cite{Bali:2011ks}
\item without and with lattice spacing corrections ($c_a a^2$ term)
\item variations of the input LECs according to the errors quoted in Table~\ref{tab:LECsuncertainties}
\end{enumerate}
Finite volume corrections are always included.

\begin{table}
  \begin{tabular}{|c||cccccc||cccccc|}
    \hline 
    & \multicolumn{6}{c||}{excluding $\sigma_{\pi N}\left(285\,\mbox{MeV}\right)$ } & \multicolumn{6}{c|}{including $\sigma_{\pi N}\left(285\,\mbox{MeV}\right)$ }\tabularnewline
    & $M_{0}$~[MeV] & $c_{1}$~[GeV$^{-1}$] & $\overline{\alpha}$~[GeV$^{-3}$] & $\frac{\chi^{2}}{dof}$ & $r_{0}$~[fm] & $\sigma_{\pi}$~[MeV] & $M_{0}$~[MeV] & $c_{1}$~[GeV$^{-1}$] & $\overline{\alpha}$~[GeV$^{-3}$] & $\frac{\chi^{2}}{dof}$ & $r_{0}$~[fm] & $\sigma_{\pi}$~[MeV]\tabularnewline
    \hline 
    $p^{2}$  & $906\left(11\right)$ & $-0.43\left(2\right)$ & -- & $2.1$ & $0.509$ & $34\left(2\right)$ & $913\left(6\right)$ & $-0.33\left(1\right)$ & -- & $6.3$ & $0.539$ & $26\left(1\right)$\tabularnewline
    $p^{3}$  & $880\left(13\right)$ & $-0.93\left(3\right)$ & -- & $1.9$ & $0.480$ & $53\left(2\right)$ & $892\left(6\right)$ & $-0.78\left(1\right)$ & -- & $8.5$ & $0.527$ & $41\left(1\right)$\tabularnewline
    $p_{\Delta}^{3}$  & $863\left(16\right)$ & $-1.19\left(4\right)$ & -- & $2.1$ & $0.456$ & $68\left(3\right)$ & $878\left(5\right)$ & $-1.00\left(1\right)$ & -- & $9.5$ & $0.517$ & $52\left(1\right)$\tabularnewline
    $p^{4}$  & $866\left(40\right)$ & $-1.18\left(14\right)$ & $23\left(3\right)$ & $2.5$ & $0.470$ & $62\left(13\right)$ & $888\left(9\right)$ & $-0.91\left(4\right)$ & $38\left(2\right)$ & $2.9$ & $0.507$ & $41\left(3\right)$\tabularnewline
    $p_{\Delta}^{4}$  & $893\left(29\right)$ & $-0.77\left(9\right)$ & $35\left(2\right)$ & $2.4$ & $0.494$ & $38\left(10\right)$ & $890\left(7\right)$ & $-0.80\left(1\right)$ & $33\left(2\right)$ & $2.5$ & $0.489$ & $41\left(2\right)$\tabularnewline
    \hline 
  \end{tabular}

  \caption{\label{tab:nf2 Fits 1)}Results for B$\chi$PT fits to $N_{f}=2$
    nucleon mass data from Refs.~\cite{Capitani:2012gj,Alexandrou:2010hf,Engel:2010my,Bali:2012qs}.
    The '$\Delta$' index denotes the inclusion of explicit $\Delta$-isobar ($\Delta$B$\chi$PT), 
    while its omission corresponds to $\s\Delta$B$\chi$PT;
    FV corrections are included but finite-spacing effects are excluded.
    The left-panel results come from a fit of solely nucleon-mass data while in the right panel
    the $\sigma_{\pi N}$ point at $M_{\pi}=285$ MeV of Ref~\cite{Bali:2011ks} was also taken into account.}
\end{table}

\begin{figure}
  \includegraphics[scale=0.43]{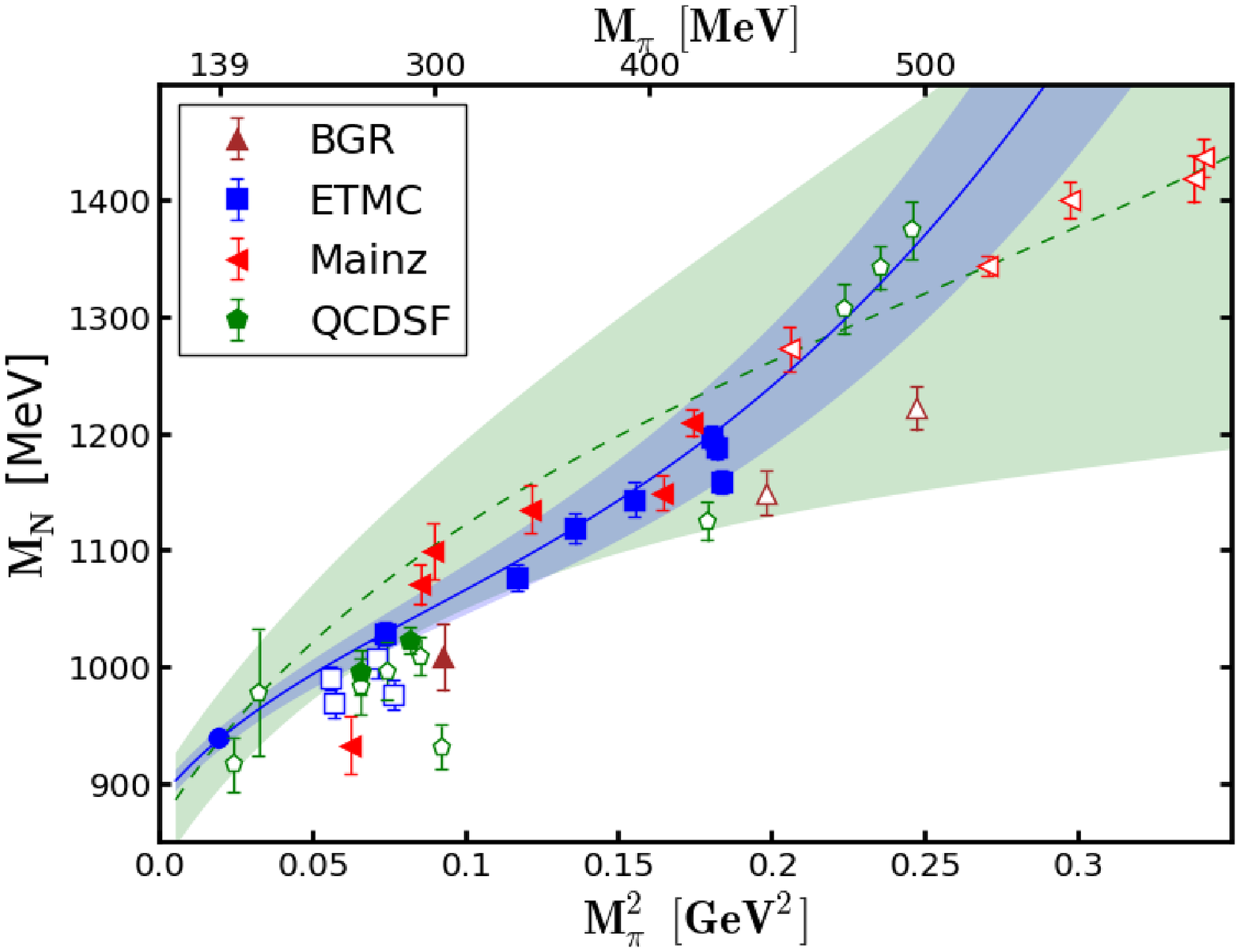}\includegraphics[scale=0.43]{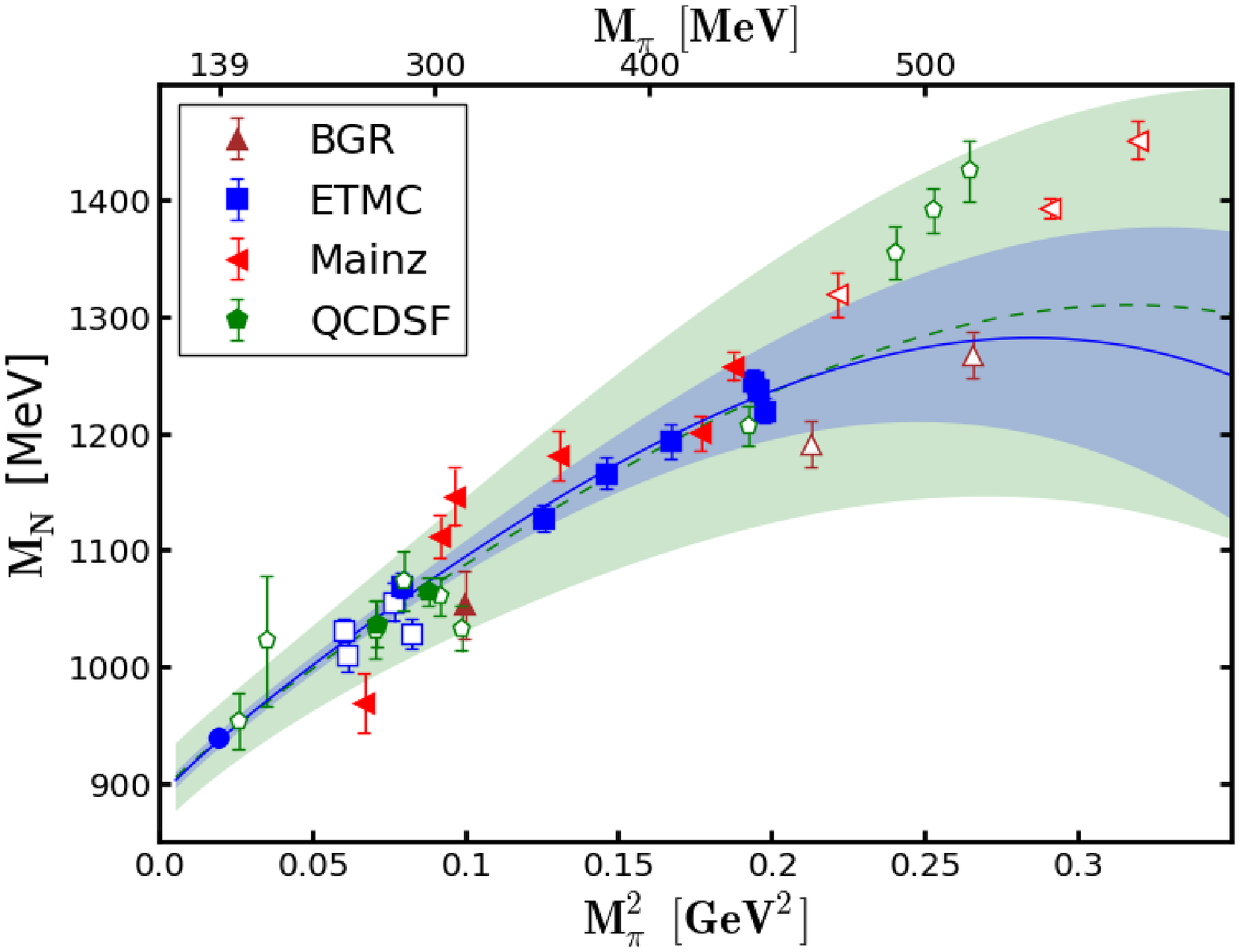}
  \caption{\label{fig:nf2 Fits 1)} Fits to the $N_f=2$
    nucleon mass data of Refs.~\cite{Capitani:2012gj,Alexandrou:2010hf,Engel:2010my,Bali:2012qs}. Filled (open) symbols are for 
data points included in (excluded from) the fits. The left (right) picture shows fits without (with) explicit $\Delta$-isobar. 
The fit including the $\sigma_{\pi N}\left(285\,\mbox{MeV}\right)$ of Ref.~\cite{Bali:2011ks} is given by the blue solid line while the 
plain nucleon mass fit is given by the green dashed one. The dark blue and light green shaded regions represent the corresponding statistical uncertainties. The lQCD data are scaled by $r_{0}$ and FV corrected according to the simultaneous fit. Hence, the green dashed line does not correspond to the shown data points.}
\end{figure}

\begin{figure}
  \begin{minipage}[t]{0.45\columnwidth}
    \includegraphics[scale=0.40]{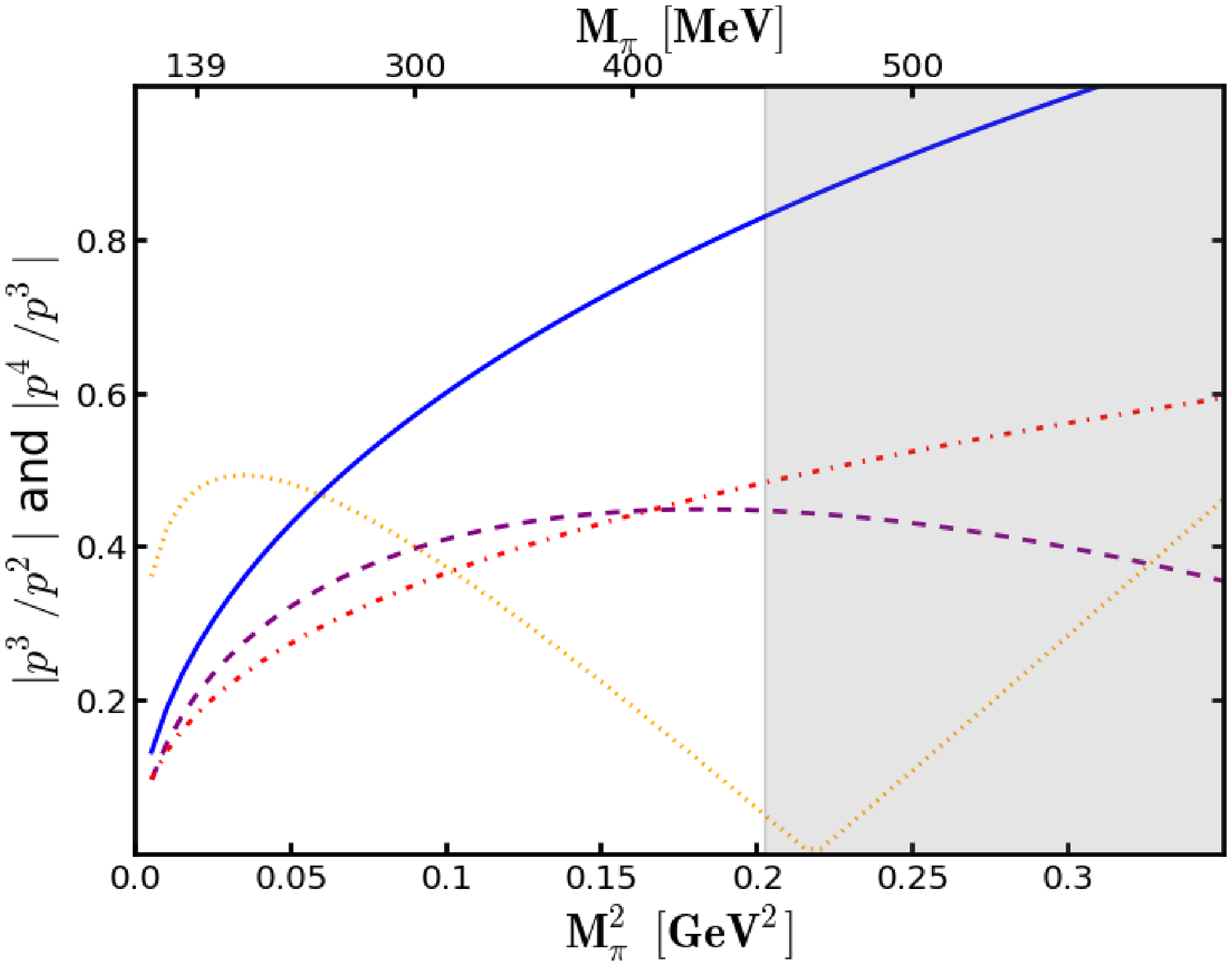}
    \caption{\label{fig:nf2 decomposed}
      The  B$\chi$PT results for $M_N(M_\pi)$ decomposed
      into their chiral-order relative contributions $|p^{3}/p^{2}|$ and $|p^{4}/p^{3}|$.
      The blue solid line denotes $|p^{3}/p^{2}|$ and the purple-dashed line, $|p^{4}/p^{3}|$, both for  
$\Delta$B$\chi$PT. The red dashed-dotted and orange dotted are the $|p^{3}/p^{2}|$ and  $|p^{4}/p^{3}|$ results for $\s\Delta$B$\chi$PT. 
 The shaded region is excluded from the fit.}
  \end{minipage}
  \hspace{0.05\columnwidth}
  \begin{minipage}[t]{0.45\columnwidth}
  \includegraphics[scale=0.40]{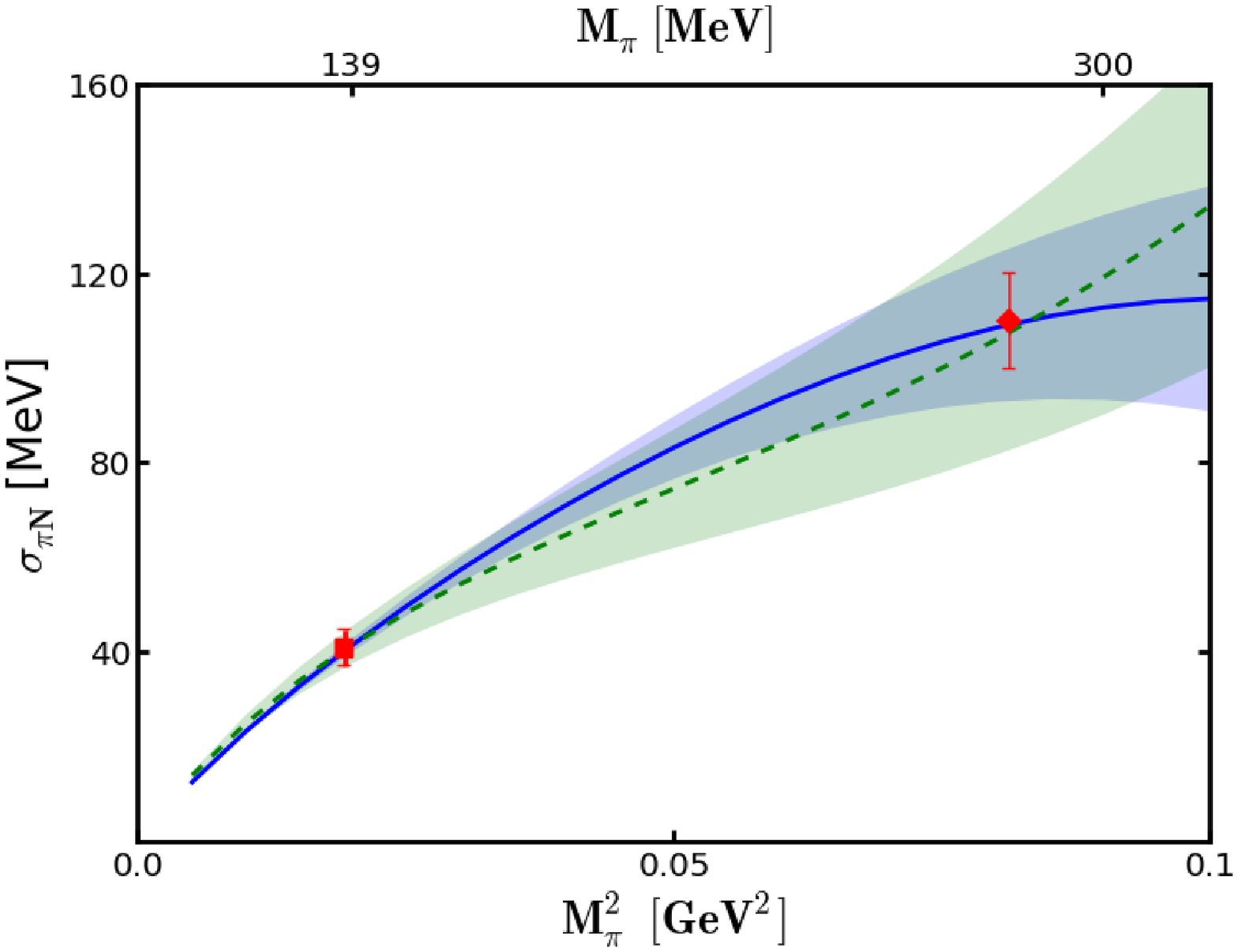}
  
  \caption{\label{fig:nf2 sigmaPiN}
    Pion-mass dependence of the $\sigma_{\pi N}$-term. The blue solid line and the green dashed lines stand for the $\Delta$B$\chi$PT and $\s\Delta$B$\chi$PT fits respectively, both including $\sigma_{\pi N}\left(285\right)$ of Ref~\cite{Bali:2011ks} (red diamond). The dark blue and light green shaded areas represent the corresponding uncertainties. The red square is our final result at the physical point.}
  \end{minipage}
\end{figure}

The output of our fits for cases 1-3, with the LECs fixed to the values in Table~\ref{tab:LECsuncertainties} and
without lattice-spacing corrections are presented in Table~\ref{tab:nf2 Fits 1)} and Fig.~\ref{fig:nf2 Fits 1)}. 
Bear in mind that changes in the fit conditions 1 and 2 yield different $r_{0}$ (see Table~\ref{tab:nf2 Fits 1)}) 
so lQCD data are scaled differently. From Table~\ref{tab:nf2 Fits 1)} we observe that the inclusion of $\mathcal{O} (p^4)$ 
does not lead to a better description of present nucleon mass data than the $\mathcal{O} (p^3)$ one. 
However, for fits including the $\sigma_{\pi N}\left(285\right)$ point, a good  $\chi^{2}/dof$ emerges only at 
$\mathcal{O} (p^4)$. In this situation, $\Delta$B$\chi$PT gives a slightly better $\chi^{2}/dof$ than $\s\Delta$B$\chi$PT but
both approaches give the same $\sigma_{\pi N}$ value. The overall rather high $\chi^{2}/dof$ is caused by two points from the 
Mainz Collaboration. By excluding them we obtain $\chi^{2}/dof \sim 1.6$ but the results change only within the quoted uncertainties. 
The FV corrections shift the data points by $(-6)-(-50)$~MeV. In contrast to the $\s \Delta$B$\chi$PT case, the $\Delta$B$\chi$PT $p^4$-results are not significantly altered by the inclusion of 
 $\sigma_{\pi N}\left(285\right)$ in the fits and exhibit a softer $M_\pi$ dependence. This might be interpreted as 
an indication that the theory with explicit $\Delta(1232)$ is more realistic.

Figure~\ref{fig:nf2 decomposed} shows the relative contributions, $|p^{3}/p^{2}|$ and $|p^{4}/p^{3}|$, 
of different chiral orders to the nucleon mass for fits including $\sigma_{\pi N}\left(285\right)$. One observes that the $\mathcal{O} (p^4)$ term has a relatively small contribution over a large $M_\pi$ range. The same is true for the $\s\Delta$B$\chi$PT $\mathcal{O} (p^3)$ term. In the $\Delta$B$\chi$PT case, however, the relative impact of the $\mathcal{O} (p^3)$ contribution steadily rises, becoming more than $80$\% of the $p^{2}$ one at $M_\pi > 450$ MeV. From this we deduce
that $M_\pi \sim 450$ MeV is at the upper border of the $\Delta$B$\chi$PT applicability. 
We have also performed fits with relaxed conditions $LM_{\pi}\geq3.5$
and $r_{0}M_{\pi}\leq1.00$ which, however, yield equivalent results
to those already presented in Table~\ref{tab:nf2 Fits 1)}. The present data
do not allow us to go below $r_{0}M_{\pi}\leq1.00$.

\begin{table}
  \begin{tabular}{|c||cccccccc|}
    \hline 
    $p_{\Delta}^{4}$  & $M_{0}$~[MeV] & $c_{1}$~[GeV$^{-1}$] & $\overline{\alpha}$~[GeV$^{-3}$] & $c_{E}$~[GeV$^{-3}$] & $c_{MQ}$~[GeV$^{-3}$] & $\frac{\chi^{2}}{dof}$ & $r_{0}$~[fm] & $\sigma_{\pi}$~[MeV]\tabularnewline
    \hline 
    excluding $\sigma_{\pi N}\left(285\,\mbox{MeV}\right)$ & $894\left(28\right)$ & $-0.76\left(10\right)$ & $36\left(5\right)$ & $-0.06\left(7\right)$ & $-0.05\left(13\right)$ & $2.8$ & $0.501$ & $37\left(10\right)$\tabularnewline
    including $\sigma_{\pi N}\left(285\,\mbox{MeV}\right)$ & $892\left(21\right)$ & $-0.79\left(2\right)$ & $34\left(3\right)$ & $-0.08\left(6\right)$ & $-0.08\left(12\right)$ & $2.8$ & $0.499$ & $40\left(3\right)$\tabularnewline
    \hline 
  \end{tabular}
  
  \caption{\label{tab:nf2 Fits 2)}Results for $p^{4}-$$\Delta$B$\chi$PT fits
    to $N_{f}=2$ nucleon mass data from Refs.~\cite{Capitani:2012gj,Alexandrou:2010hf,Bali:2012qs} with lattice spacing effects 
accounted by the $c_{E}a^{2}$ and $c_{MQ}a^{2}$ terms for the  ETMC and Mainz/QCDSF data respectively.}
\end{table}

In Table~\ref{tab:nf2 Fits 2)} we summarize our results including finite-lattice spacing corrections in the fit, namely the 
$c_{E}a^{2}$ and $c_{MQ}a^{2}$ terms for ETMC and Mainz/QCDSF respectively. We obtain corrections of
$(+6)-(+20)$~MeV, which have an opposite sign with respect to the FV corrections. 
By comparing to Table~\ref{tab:nf2 Fits 1)}
we notice that all changes are within the already given uncertainties.
A noticeable qualitative effect is that changes in the Sommer-scale counterbalance 
finite-lattice spacing corrections so that the results are close to the former ones.
A more elaborated EFT background is required to calculate and interpret finite-lattice spacing corrections more reliably.

We have tested the fits for variations of $c_{2}$, $c_{3}$ within the
errors given in Table~\ref{tab:LECsuncertainties}. In all cases the results
are compatible within uncertainties with those of Table~\ref{tab:nf2 Fits 1)}.
We conclude that the $p^{4}$ B$\chi$PT fits are not able to constrain
these LECs effectively.

Furthermore, by varying $c_{1\Delta}$ we find it to be correlated with $\overline{\alpha}$. 
The inclusion of $c_{1\Delta}$ as a free parameter does not produce sensible fits unless 
the $\sigma_{\pi N}\left(285\right)$ point is taken into account. The fit is driven to 
unreasonable high $c_{1\Delta}$ with rather large $\overline{\alpha}$ values. However, in 
fits including the $\sigma_{\pi N}\left(285\right)$ point we recover 
$c_{1\Delta}=-0.87\left(16\right)$ GeV$^{-1}$ together with results 
compatible with those in Table~\ref{tab:nf2 Fits 1)}. A scan over a range of
$c_{1\Delta}$ shows that reasonable fits can only be obtained for the 
interval $c_{1\Delta}=(-0.8)-(-1.0)$ GeV$^{-1}$, resulting in $\sigma_{\pi N}$ 
values in the range $37$-$45$ MeV. We observe that the correlation between $c_{1\Delta}$ and 
$\overline{\alpha}$ is relaxed by the addition of the $\sigma_{\pi N}\left(285\right)$ point. 

As a final $\sigma_{\pi N}$-value for the $N_{f}=2$ lQCD fits we quote
\[
\sigma_{\pi N}=41\left(5\right)\left(4\right)\,\,\,\mbox{MeV},
\]
which corresponds to our $p^{4}$ $\s\Delta$ and $\Delta$B$\chi$PT
fits of Table \ref{tab:nf2 Fits 1)}
including $\sigma_{\pi N}\left(285\right)$ and FV corrections.
The first uncertainty is statistical and can be taken, as a first approximation, 
to be 3~MeV, which is the largest error from the fits under consideration. However, 
one should note that we obtain $\chi^{2}/d.o.f. > 1$, that we interpret as an indication 
of underestimated uncertainties in the data. To correct for this, we repeat the 
fits  multiplying the statistical errors of all points by $\sqrt{\chi^{2}/d.o.f.}$, 
in analogy to the procedure adopted by the Particle Data Group~\cite{Beringer:1900zz} 
for unconstrained averages. The new error of 5 MeV is the largest one, corresponding 
to the $\s\Delta$B$\chi$PT case. The systematic uncertainty, second figure, 
is determined by adding in quadratures the variation induced by changes in $c_{1\Delta}$ 
in the range given above to the finite spacing effects (Table~\ref{tab:nf2 Fits 2)}). 
In an attempt to identify any additional bias in the data samples, 
we have performed new fits using the {\it delete-1 jackknife} technique. 
The resulting fit values and errors did not differ significantly from the quoted ones. 
Note that the single $\sigma_{\pi N}\left(285\right)$ measurement has a strong influence 
on our $N_f =2$ result. Indeed by excluding this point and averaging over the $\Delta$B$\chi$PT and $\s \Delta$B$\chi$PT results  we get a 
$\sigma_{\pi N}=52\left(13\right)\left(11\right)$ MeV, albeit with large errorbars. 
In view of this, new direct $\sigma_{\pi N}$ measurements at low pion masses will be important 
to establish the actual value of this quantity.

Figure~\ref{fig:nf2 sigmaPiN} summarizes our results for the pion mass dependence of 
the $\sigma_{\pi N}$-term. The results for the $\Delta$B$\chi$PT and $\s\Delta$B$\chi$PT fits are compatible within errors but exhibit 
a different $M_\pi$ dependence.

For our final values of the LECs $M_0$, $c_1$ and $\overline{\alpha}$ we quote those of the 
$p^4$-$\Delta$B$\chi$PT fit of Tab. \ref{tab:nf2 Fits 1)} including $\sigma(285)$. In particular, in the present work we set 
the Sommer-scale to $r_0=0.493(23)$~fm, which is the average of all our $p^4$ results and where the uncertainty is chosen such as to cover all our $p^4$ fits.  

\subsection{Nucleon mass up to order $\mathcal{O}\left(p^{4}\right)$: fits to
  $N_{f}=2+1$ lattice QCD data \label{sub:Nf21 fits}}

We use our B$\chi$PT nucleon mass formula of Eq. (\ref{eq:MN(4)(Mpi)}) to fit the lQCD data for the $N_{f}=2+1$ ensembles of different collaborations with $M_{\pi} L >3.8$ and $M_{\pi} \lesssim 415$ MeV. Thus, we include 9 points from the BMW collaboration~\cite{Durr:2011mp}, 1 point from HSC~\cite{Lin:2008pr}, 1 from LHPC~\cite{WalkerLoud:2008bp,Bratt:2010jn}, 4 fine and 4 super-fine from MILC~\cite{Aubin:2004wf,MILC:2013}, 3 from NPLQCD~\cite{Beane:2011pc}, 2 from PACS-CS~\cite{Aoki:2008sm} and 6 from RBC-UKQCD~\cite{Jung:2012rz}. The selected data have already been corrected to the physical strange quark mass (BMW) or come from configurations for which the strange quark mass (in the {$\overline{MS}$ scheme at 2~GeV) has been reported to be close enough to the physical limit, to make the corresponding correction negligible~\footnote{Notice that the small strange quark mass found in Ref.~\cite{Aoki:2008sm}, $m_s^{\overline{MS}} \sim 72$ MeV, has been attributed to the perturbative approach employed in that paper to relate lattice- and the $\overline{MS}$-renormalized values~\cite{Aoki:2009ix}.}. The approach of the QCDSF-UKQCD collaboration~\cite{Bietenholz:2011qq,Horsley:2011wr} is conceptually different as it generates points along the $SU\left(3\right)$ singlet line, $2\overline{m}+m_{s}={\mathrm const}$. Therefore in these simulations both the light and strange quark masses remain unphysical, making our $SU\left(2\right)$ approach not applicable.

Most of the data are provided in terms of $\left(aM_{\pi},aM_{N}\right)$, 
together with the individual lattice spacings $a$ and the statistical uncertainties for all the three quantities.
Unlike the $N_f =2$ case, the available information does not allow us to perform our own scale setting. 
Therefore, we treat the $a$-uncertainties as correlated normalization errors for all $M_{N}$ points from the same set. 
Our treatment of normalization uncertainties follows from Ref.~\cite{D'Agostini:1995fv}.
We perform three types of fits: 1) neglecting correlated normalization errors, 2) including the normalization error in scale factors $f_{i}$, 3) including the normalization uncertainty in a correlation matrix $V$.  For the case 3) we also consider lattice spacing effects. The $\chi^{2}$ functions for type 2 and 3 fits read
\begin{eqnarray}
\label{eq:chi1}
  \chi_{2}^{2} & = & \sum_{i}\left[\frac{M_{N}^{(n)}\left(M_{\pi}^{2}\right)+\Sigma_{N}^{(n)}\left(M_{\pi}^{2},L\right)-f_{i}d_{i}\left(M_{\pi}^{2},L\right)}{f_{i}\sigma_{i}}\right]^{2}+\left[\frac{f_{i}-1}{\sigma_{f_{i}}}\right]^{2}\,\,\,\,,\\
  \chi_{3}^{2} & = & \vec{\Delta}^{T}V^{-1}\vec{\Delta}\,\,\,\,\,\,\,\,\,\,\,\,\mbox{with}\,\,\,\,\,\,\,\,\,\,\,\,\Delta_{i}=\left[M_{N}^{(n)}\left(M_{\pi,i}^{2}\right)+c_{i}a_{i}^{2}+\Sigma_{N}^{(n)}\left(M_{\pi,i}^{2},L_{i}\right)-d_{i}\left(M_{\pi,i}^{2},L_{i}\right)\right]\,\,\,\,,
\end{eqnarray}
where $M_{N}^{(n)}\left(M_{\pi}^{2}\right)$ and $\Sigma_{N}^{(n)}\left(M_{\pi}^{2},L\right)$
are the B$\chi$PT continuum and finite volume expressions given in App. \ref{Appendix sub:Fit-formulas}; $d_{i}\left(M_{\pi}^{2},L\right)$ are the lQCD data, each point for a given lattice size $L$ and spacing $a$.
We denote the statistical uncertainty for $M_{N}$ coming from $aM_{N}$
as $\sigma_{i}$ and the normalization uncertainty coming from $a$
as $\sigma_{f_{i}}$. Case 1) is recovered from Eq.~(\ref{eq:chi1}) by taking all $f_{i}=1$ and replacing  $\sigma_i \rightarrow \sqrt{\sigma_{i}^{2}+\sigma_{f_{i}}^{2}}$ corresponding to the assumption that $\sigma_{i}$ and $\sigma_{f_{i}}$ are uncorrelated errors. In case 2) the $f_{i}$ are additional fit parameters; $\sigma_{i}$ 
and $\sigma_{f_{i}}$ are treated separately. In case 3)
$\sigma_{i}$ and $\sigma_{f_{i}}$ are incorporated in the
correlation matrix $V$. The BMW collaboration \cite{Durr:2011mp}
does not provide enough information to disentangle the uncertainties from
$aM_{N}$ and $a$ so that we always include this data set with uncorrelated
uncertainties.

In our fits,  the LECs $c_{2}$, $c_{3}$ and $c_{1\Delta}$ are fixed to the values given in Table~\ref{tab:LECsuncertainties}. 
There are two points with  $M_\pi \sim 390$ MeV from Refs.~\cite{Lin:2008pr,Beane:2011pc} with very small 
reported $\sigma_{i}$ and slightly smaller $M_N$ values compared to the neighboring points (see Fig. \ref{fig:Fits nf21}). 
The inclusion of these points shifts the results to lower masses, yielding a slightly worse $\chi^2/dof$. 
Although these points were obtained by different NPLQCD and HSC Collaborations, they are not entirely independent because 
NPLQCD uses the scale of the HSC Collaboration, which actually expresses some concern about the quality of their lattice-spacing 
determination. In view of the situation, we exclude these two points from our main results but consider their influence in the 
systematic uncertainties.
 
\begin{figure}
  \includegraphics[scale=0.43]{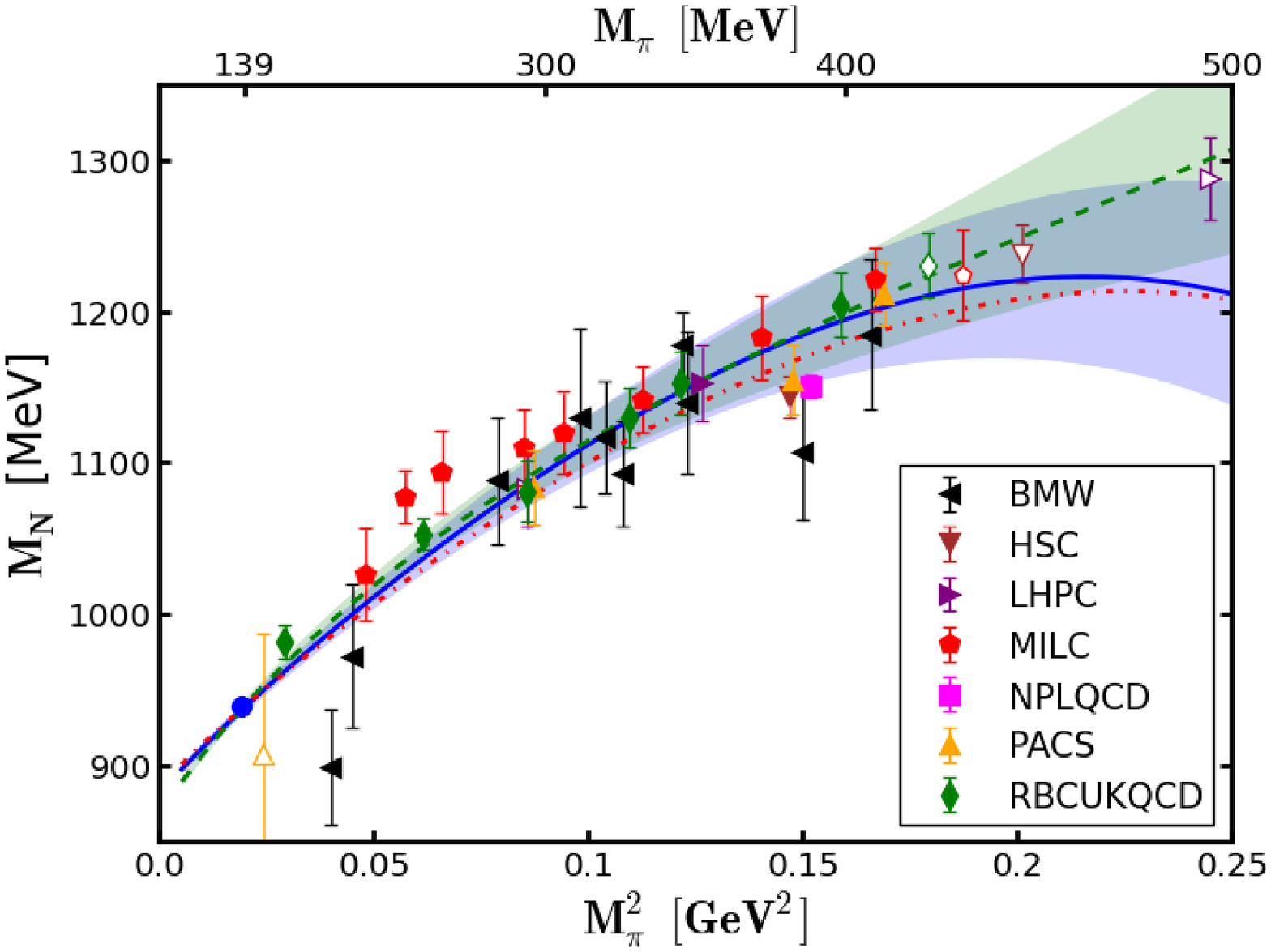}
  \includegraphics[scale=0.43]{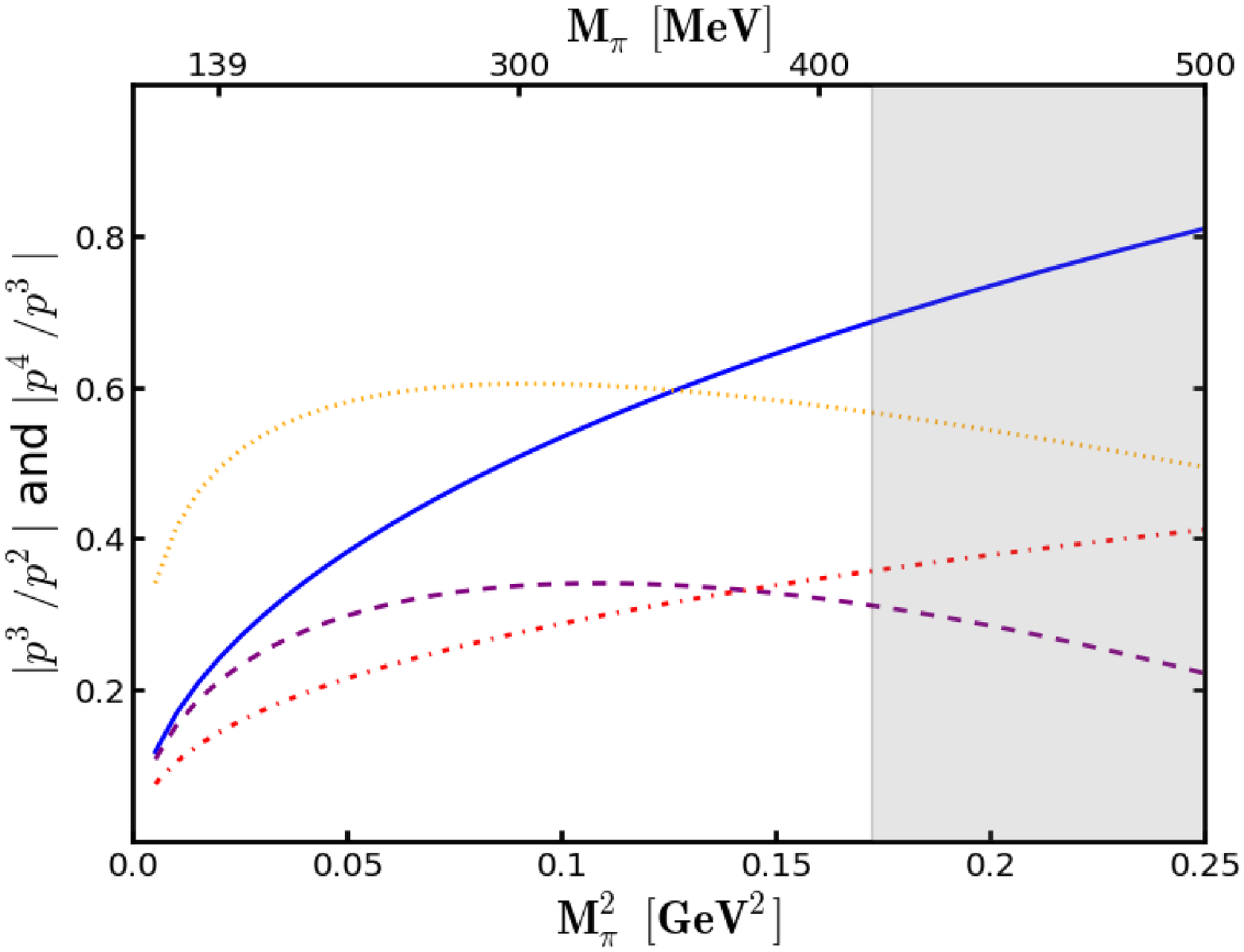}
  
  \caption{\label{fig:Fits nf21} Combined fits to lQCD data of the $N_{f}=2+1$ ensembles
    \cite{Aoki:2008sm,Durr:2011mp,Bratt:2010jn,Jung:2012rz,MILC:2013,Lin:2008pr,Beane:2011pc}.
    Left: Fits to nucleon mass data up to $M_{\pi}=415$ MeV. The blue solid (green dashed) line 
shows the $\mathcal{O}(p^4)$ $\Delta$B$\chi$PT ($\s \Delta$B$\chi$PT) fit of type 3). 
The red dotted line is also for $\mathcal{O}(p^4)$ $\Delta$B$\chi$PT but including the two points 
of  $M_\pi\sim 390$ MeV, excluded from the main fits as discussed in the text.
Filled (open) symbols represent points included in (excluded from) the fits.
Right: Decomposition of the fit results in their chiral order contributions. 
The blue solid line corresponds to the $|p^{3}/p^{2}|$ ratio and the purple-dashed one to $|p^{4}/p^{3}|$, 
both for $\Delta$B$\chi$PT. The red dashed-dotted and orange-dotted are the  $|p^{3}/p^{2}|$ and  $|p^{4}/p^{3}|$ 
results obtained with $\s\Delta$B$\chi$PT.}
\end{figure}

In Table~\ref{tab:nf21 fits} we display our results for 
the fit types 1) and 3). The results obtained with option 2)  are similar 
to those obtained with 3) so we do not show them. The consideration of 
normalization uncertainties slightly enhances the $\chi^{2}/dof$ 
but causes a noticeable reduction of $c_1$ and $\sigma_{\pi N}$. 
The quality of the fits in terms of $\chi^{2}/dof$ is essentially
the same for $p^{3}$ and $p^{4}$ fits. As in the $N_f = 2$ case, we expect 
the advantage of the $p^{4}$ formula to be tangible as soon as 
direct $\sigma_{\pi N}$-data for low pion masses become available for $N_{f}=2+1$. 

The left panel of Fig.~\ref{fig:Fits nf21} shows the pion mass dependence of 
our $\mathcal{O}(p^4)$ nucleon mass results for both $\s \Delta$B$\chi$PT and $\Delta$B$\chi$PT. 
There is a large overlap of the 
corresponding error bands, which are addressed below in more detail.
By decomposing the fits into their chiral-order relative contributions (right panel
of Fig. \ref{fig:Fits nf21}), we observe a similar situation to the $N_f=2$ case. 
Namely, the $\mathcal{O}(p^4)$ relative contributions are small over a large range of $M_\pi$
but the $\mathcal{O}(p^3)$ in $\Delta$B$\chi$PT increases, making the applicability of our 
perturbative expression questionable for high $M_\pi$ values.
We have checked that a fit constrained to $M_{\pi} <360$~MeV produces 
results compatible with those of the $M_{\pi} < 415$~MeV fit but with larger uncertainties.

\begin{table}
  \begin{tabular}{|c||ccccc||ccccc|}
    \hline 
    $LM_{\pi}\geq3.8$ & \multicolumn{5}{c||}{$\s\Delta$B$\chi$PT} & \multicolumn{5}{c|}{$\Delta$B$\chi$PT}\tabularnewline
    $M_{\pi}\leq415$ MeV & $M_{0}$~[MeV] & $c_{1}$~[GeV$^{-1}$] & $\overline{\alpha}$~[GeV$^{-3}$] & $\frac{\chi^{2}}{dof}$ & $\sigma_{\pi}$~[MeV] & $M_{0}$~[MeV] & $c_{1}$~[GeV$^{-1}$] & $\overline{\alpha}$~[GeV$^{-3}$] & $\frac{\chi^{2}}{dof}$ & $\sigma_{\pi}$~[MeV] \tabularnewline
    \hline 
    $p^{2}$  & $904\left(2\right)$ & $-0.47\left(1\right)$ & -- & $3.1$ & $36\left(1\right)$ & -- & -- & -- & -- & --\tabularnewline
    $p^{3}$  & $883\left(2\right)$ & $-0.90\left(1\right)$ & -- & $1.3$ & $51\left(1\right)$ & $870\left(2\right)$ & $-1.10\left(1\right)$ & -- & $1.2$ & $60\left(1\right)$\tabularnewline
    $p^{4}$  & $870\left(3\right)$ & $-1.15\left(3\right)$ & $24\left(2\right)$ & $1.3$ & $58\left(3\right)$ & $883\left(3\right)$ & $-0.89\left(3\right)$ & $26\left(2\right)$ & $1.4$ & $49\left(2\right)$\tabularnewline \hline
    no correl. $p^{4}$ & $865\left(5\right)$ & $-1.22\left(5\right)$ & $19\left(4\right)$ & $1.0$ & $63\left(3\right)$ & $878\left(4\right)$ & $-0.96\left(4\right)$ & $20\left(4\right)$ & $1.1$ & $54\left(3\right)$\tabularnewline
    no correl. (390) $p^{4}$ & $863\left(5\right)$ & $-1.25\left(5\right)$ & $15\left(4\right)$ & $1.4$ & $64\left(3\right)$ & $876\left(4\right)$ & $-0.99\left(4\right)$ & $15\left(3\right)$ & $1.6$ & $56\left(3\right)$\tabularnewline

    \hline 
  \end{tabular}

  \caption{\label{tab:nf21 fits} Combined fits to the $N_{f}=2+1$ lQCD ensembles
    \cite{Aoki:2008sm,Durr:2011mp,Bratt:2010jn,Jung:2012rz,MILC:2013,Lin:2008pr,Beane:2011pc}
    for pion masses $M_{\pi}\leq415$ MeV. The LECs $c_{2}$, $c_{3}$
    and $c_{1\Delta}$ are set to the central values given in  Table~\ref{tab:LECsuncertainties};
    FV effects are included while $a^{2}$ effects are excluded. The last two rows correspond to fits of type 1) neglecting correlated normalization errors. The fit of the last row takes into account the two points of Refs.~\cite{Lin:2008pr,Beane:2011pc} with  $M_\pi\sim 390$ MeV, excluded from the main fits as discussed in the text.}
\end{table}

The results of the fits taking into account lattice spacing effects are given in Table~\ref{tab:nf21 a^2}. 
These are considered for data sets with enough points with the same $L$ and different $a$ values. Explicitly, we
introduced two terms $c_{M}a^{2}$ and $c_{R}a^{2}$ for the MILC
and RBCUK Collaborations, respectively. In the case of  BMW, we assume that lattice spacing uncertainties are included in 
the errorbars. We find nucleon mass shifts of $(-7)-(-46)$~MeV, which are small but comparable in size with the FV corrections. 
With this correction, the $\chi^{2}/dof$ is slightly better and $\sigma_{\pi N}$ gets smaller by several MeV. 
The uncertainties for the constants $c_{M}a^{2}$ and $c_{R}a^{2}$ are now slightly smaller than in the 2 flavor case although 
all values of Tables~\ref{tab:nf2 Fits 2)} and \ref{tab:nf21 a^2} agree within the individual errors.

\begin{table}
  \begin{tabular}{|c||ccccccc|}
    \hline 
    & $M_{0}$~[MeV] & $c_{1}$~[GeV$^{-1}$] & $\overline{\alpha}$~[GeV$^{-3}$] & $c_{M}$~[GeV$^{-3}$] & $c_{R}$~[GeV$^{-3}$] & $\frac{\chi^{2}}{dof}$ & $\sigma_{\pi}$~[MeV]\tabularnewline
    \hline 
    $\s\Delta$B$\chi$PT  & $873\left(4\right)$ & $-1.10\left(5\right)$ & $27\left(3\right)$ & $0.18\left(8\right)$ & $0.03\left(2\right)$ & $1.2$ & $55\left(3\right)$\tabularnewline
    $\Delta$B$\chi$PT & $887\left(3\right)$ & $-0.84\left(4\right)$ & $29\left(3\right)$ & $0.21\left(8\right)$ & $0.04\left(2\right)$ & $1.2$ & $44\left(3\right)$\tabularnewline
    \hline 
  \end{tabular}

  \caption{\label{tab:nf21 a^2} Combined fits to the $N_{f}=2+1$ lQCD ensembles
    \cite{Durr:2011mp,Jung:2012rz,MILC:2013} including $c_a a^{2}$ corrections for the MILC ($c_{M}$) and RBCUK ($c_{R}$)
    Collaborations. The LECs $c_{2}$, $c_{3}$ and $c_{1\Delta}$ are
    set to the central values in Table~\ref{tab:LECsuncertainties}.}
\end{table}

We tested our results for changes by varying $c_{2}$, $c_{3}$ and
$c_{1\Delta}$ within the errors quoted in Table \ref{tab:LECsuncertainties}. 
All changes are within the above quoted uncertainties. In particular,
changes in $c_{1\Delta}$ are compensated by changes in $\overline{\alpha}$
and reasonable results are only obtained for the 
range of $c_{1\Delta}=(-0.5)-(-1.3)$ GeV$^{-1}$ estimated above.

As a final value for $\sigma_{\pi N}$ in the $N_{f}=2+1$ case we
give 
\[
\sigma_{\pi N}=52\left(3\right)\left(8\right)\,\,\,\,\mbox{MeV}\,\,\,\,,
\]
obtained in the following way. The central value is the average of 
the four $\mathcal{O} (p^{4})$ $\Delta$B$\chi$PT and $\s\Delta$B$\chi$PT results without (Table~\ref{tab:nf21 fits}) 
and with (Table \ref{tab:nf21 a^2}) lattice spacing corrections, all including correlated normalization uncertainties. 
The first error corresponds to the largest statistical uncertainty of the values under consideration and the second 
is the largest difference among them.

\begin{figure}
  \includegraphics[scale=0.43]{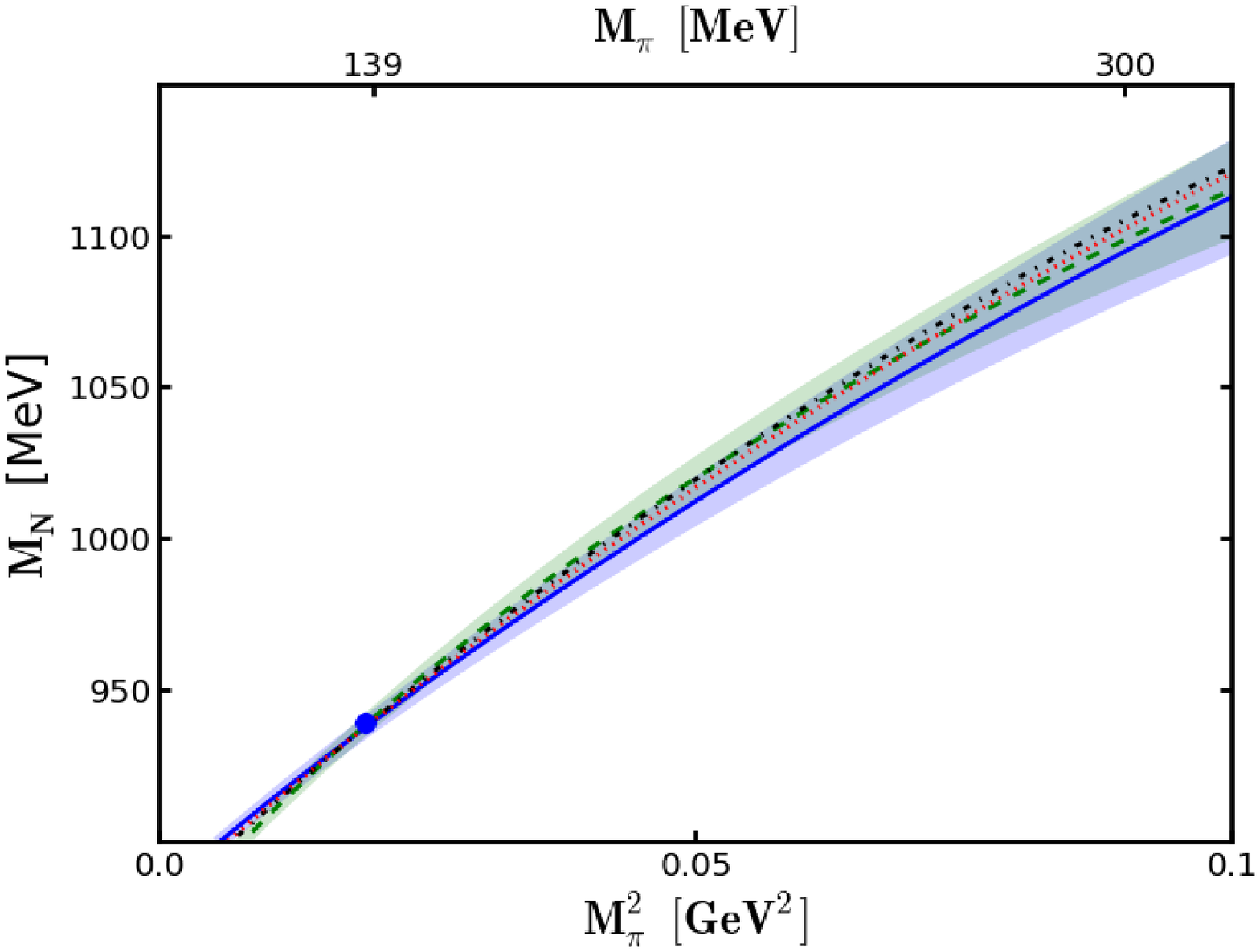}
  \includegraphics[scale=0.43]{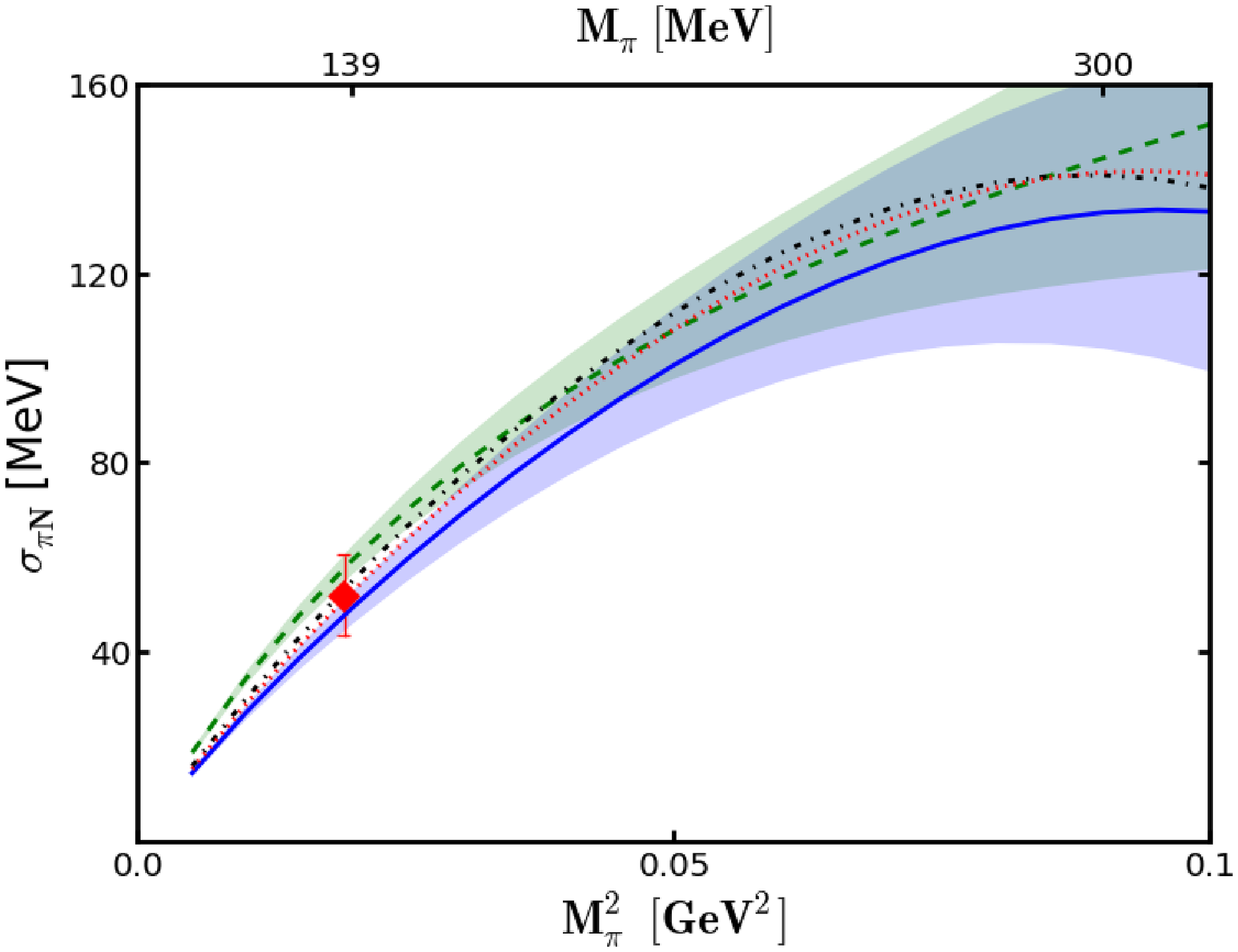}

  \caption{\label{fig:sigma piN nf21} Pion mass dependence of $M_N$
    and $\sigma_{\pi N}$ given by different $\mathcal{O} (p^{4})$ B$\chi$PT fits to $N_f=2+1$ data. The blue solid and green dashed lines 
stand for $\Delta$B$\chi$PT  and $\s\Delta$B$\chi$PT. The red dotted line is the $\Delta$B$\chi$PT solution with data points only up to $360$ MeV. The black dashed-dotted line does not take correlated normalization uncertainties into account.
The blue circle is the phenomenological nucleon mass and the red square is our $\sigma_{\pi N}$ result at the physical point. }
\end{figure}

\begin{figure}
  \includegraphics[scale=0.43]{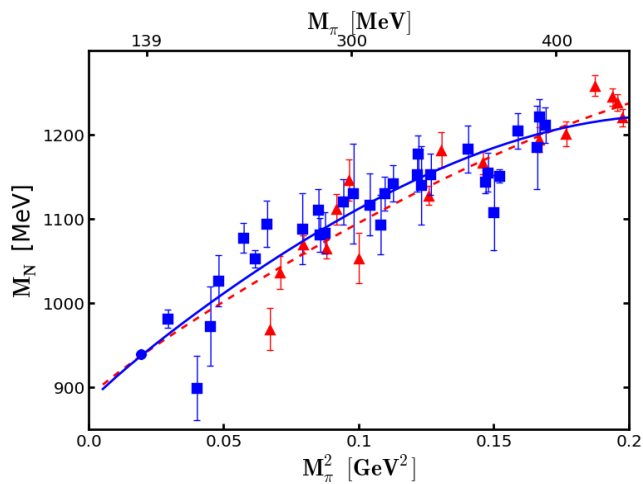}

  \caption{\label{fig:nf2 and nf21 comp} Pion mass dependence of the nucleon mass. The blue solid line and blue squares correspond to our fits to $N_f=2+1$ lQCD data. The red-dashed line and red-triangles correspond to our fits to $N_f=2$ lQCD data including the $\sigma_{\pi N}(285)$ point. The errorbands for our fit results have been removed for
the sake of clarity.}
\end{figure}

Further conclusions can be extracted from Fig.~\ref{fig:sigma piN nf21} where the pion mass dependence of $M_N$ and $\sigma_{\pi N}$ 
is shown for various $p^4$ fit strategies.  We can see that the small slope variations in $M_N(M_\pi)$ (left plot) translate into changes in 
$\sigma_{\pi N}$ of less than 10 MeV at the physical point (right plot). One also notices that the uncertainties of the
individual lQCD data points (see Fig.~\ref{tab:nf21 a^2}) tend to be larger than these variations. We do not
expect that with more low-$M_\pi$ nucleon mass data points one would be able to reduce the $\sigma_{\pi N}$ uncertainty much further, 
although simulations using one lattice action and different lattice spacings would be very important for a systematic treatment of 
discretization uncertainties. On the other hand, $N_f = 2+1$ direct measurements
of $\sigma_{\pi N}$ at low $M_\pi \lesssim 300$ MeV would probably lead to better constrained fits as it happens for $N_f =2$, reducing 
 uncertainties significantly.

Another outcome of our analysis is a slight disagreement between
the determinations of $\sigma_{\pi N}$ using either $N_{f}=2$ or
$2+1$ data. The lQCD data available at present do not allow to 
establish unambiguously the origin of this discrepancy. First of all, 
it is instructive to compare the $\Delta$B$\chi$PT $p^{4}$ fits given 
in Tables~\ref{tab:nf2 Fits 1)} and \ref{tab:nf21 fits} (also shown in 
Fig.~\ref{fig:nf2 and nf21 comp}). The corresponding $\sigma_{\pi N}$ 
values decomposed in their chiral $p^{2}$, $p^{3}$ and $p^{4}$ contributions 
are $41$ MeV$=62-27+6$ MeV and $49$ MeV$=69-26+6$ MeV, respectively. Most of 
the difference comes from the $p^{2}$ term, which is more effectively constrained 
by data points in the low $M_{\pi}$ region. New $N_{f}=2$ measurements in this 
region might help to understand the origin of the difference. On the other hand, 
a closer look to Tables~\ref{tab:nf2 Fits 2)} and \ref{tab:nf21 a^2} reveals that 
the $N_{f}=2$ and $2+1$ $\Delta$B$\chi$PT $p^{4}$ results become consistent once 
finite spacing corrections are considered. However, while the differences between 
$\s\Delta$ and $\Delta$B$\chi$PT disappear in $N_{f}=2$ after the 
$\sigma_{\pi N}\left(285\right)$ point is included in the fits, they remain in the 
$N_{f}=2+1$ case, where such a direct measurement is not available. Future direct 
determinations of $\sigma_{\pi N}$ at low pion masses for both $N_{f}=2$ or
$2+1$ data will be crucial to discriminate between different theoretical descriptions 
and to establish the value of $\sigma_{\pi N}$ at the physical point with high precision. 
Finally, we cannot exclude that part of the observed discrepancy arises from the different 
role played by strange quarks in $N_{f}=2$ simulations where they are quenched, and in 
$N_{f}=2+1$ ones, where they are dynamical and more realistic. In conclusion, 
we think our analysis exploits the considerable size of the current data set on $M_{N}$
in a way that it is possible to become sensitive to unexpected systematic
effects. However, more lQCD data will be required to settle this issue
and interpret possible discrepancies of this type.

\section{Summary and conclusion}
\label{sec:conclusions}

We have studied the nucleon mass and the $\sigma_{\pi N}$-term
in the $SU\left(2\right)$ covariant B$\chi$PT up to the chiral order $p^{4}$.
We have performed fits, using B$\chi$PT with and without explicit $\Delta$-isobar
degrees of freedom, to combined lQCD data from various Collaborations
for  $N_{f}=2$ and $N_{f}=2+1$ numbers of flavors. Special attention has been   
payed to the different sources of uncertainties in the input data. 
This study is the first application of the $p^4$ $SU\left(2\right)$ covariant B$\chi$PT
with the EOMS renormalization scheme and consistent treatment of the $\Delta$-isobar to lQCD data.
We have included finite volume corrections and also discussed finite spacing effects.
In the $N_{f}=2$ case we were able to set the lQCD data normalization via the Sommer-scale $r_0$ and 
also performed simultaneous fits to nucleon mass data and one available low $M_\pi$ $\sigma_{\pi N}$ data point.
In the $N_{f}=2+1$ case we took into account correlated normalization uncertainties for points belonging
to the same data set. In the following we summarize our findings.

\begin{itemize}
\item Our formula for the nucleon mass depends on several low energy constants, some of which have been fitted to the lQCD data.  Explicitly, the LECs are $M_{0}$, $c_{1}$, $c_{2}$, 
  $c_{3}$, $c_{1\Delta}$, $M_{\Delta0}$,
  $g_{A}$, $f_{\pi}$, $h_{A}$ and $\overline{\alpha}$; the latter is a linear combination of several couplings that appear in the chiral Lagrangian at $\mathcal{O}(p^{4})$. We adopted the phenomenological
  values for $g_{A}$, $f_{\pi}$ and $h_{A}$. Our fits are 
  insensitive to the chosen values of $c_{2}$, $c_{3}$, $c_{1\Delta}$
  and $M_{\Delta0}$ so that we are not able to constrain $c_{2}$ and $c_{3}$ and fix them to phenomenological values extracted from $\pi N$-scattering.
Furthermore, we observe that 
  $c_{1\Delta}$ and $\overline{\alpha}$ are correlated, which hinders 
  a better determination of $c_{1\Delta}$ than the range $c_{1\Delta}=(-0.5)-(-1.3)$~GeV$^{-1}$ based on rather scarce lQCD data for the $\Delta(1232)$ mass. The LECs $M_{0}$, $c_{1}$ and 
  $\overline{\alpha}$ are better determined, and their values are listed
  in Tables \ref{tab:nf2 Fits 1)} and \ref{tab:nf21 fits} for the $N_{f}=2$
  and $N_{f}=2+1$ fits. 
  For the $N_{f}=2$ ensembles we were able to 
  extract the Sommer-scale, finding $r_{0}=0.493\left(23\right)$ fm. 
By performing  fits to nucleon mass data alone
  as well as including a $\sigma_{\pi N}$ lQCD data point
  at $M_{\pi}=285$ MeV from the QCDSF Collaboration we have obtained that the inclusion of 
the  $p^{4}$ order improves the quality of the simultaneous fits.

\item For both $N_f =2$ and $2+1$ ensembles we have investigated the 
effects coming from finite
  lattice spacings $a$ and volumes employed in lQCD.  We parametrized lattice-spacing effects 
  by linear $a^{2}$ terms and applied the standard
  B$\chi$PT FV corrections.  We have obtained that both effects yield comparable numerical corrections 
  to the nucleon mass. However, we also found that the simple parametrization of the finite lattice spacing 
  effects does not allow to disentangle it in a quantitative
  manner from other effects. Fit results with and without finite $a^2$-effects are compatible within
  the statistical uncertainty. In contrast to the $a^2$-effects, the FV corrections are much better
  under control due to the established B$\chi$PT techniques for the presently available lQCD volumes.

\item We have extracted the $\sigma_{\pi N}$-term for the $N_{f}=2$ and $N_{f}=2+1$
  lQCD ensembles obtaining $\sigma_{\pi N}=41\left(5\right)\left(4\right)$
  MeV and $\sigma_{\pi N}=52\left(3\right)\left(8\right)$, respectively.
  The inclusion of the $N_f =2$ $\sigma_{\pi N}$ data point greatly reduces the 
  $\sigma_{\pi N}$ uncertainty as well as brings the two approaches, 
  $\Delta$B$\chi$PT and $\s\Delta$B$\chi$PT, closer. 
  In the case of the $N_{f}=2+1$ ensembles, where we fitted solely nucleon mass data, 
  the two approaches give $\sigma_{\pi N}$-values that differ by $9$ MeV. 
  This is a novel feature with respect to 
  HB$\chi$PT fits where the inclusion of the $\Delta$-isobar alters the result by 
   more than $40$ MeV~\cite{WalkerLoud:2008pj}. 
   The inclusion of finite lattice spacing correction to the $N_{f}=2+1$ data tends to 
  reduce $\sigma_{\pi N}$. 
  Furthermore, we want to call the attention to the fact that our result in $N_f=2$ is only 
  compatible with the experimental determination based on the KA85 $\pi N$ scattering partial 
  wave analyses of Refs~\cite{Gasser:1990ce,Alarcon:2011zs}. Our $N_f = 2+1$ value is also 
  compatible with the latest determination from the WI08 and EM06 analyses, $\sigma_{\pi N} = 59 (7)$, 
  which is phenomenologically favored on the grounds of consistency with $\pi N$ 
  phenomenology~\cite{Alarcon:2011zs}. Finally, this $N_f = 2+1$ result would lead, according to 
  the traditional arguments linking sigma terms to the baryon-octet mass 
  splittings~\cite{Gasser:1980sb,Borasoy:1996bx}, to a large strangeness content 
  in the nucleon. However, the uncertainties in these arguments have been recently 
  revisited~\cite{Alarcon:2012nr} with the conclusion that a $\sigma_{\pi N}$ of this 
  size is not at odds with, but favored by a negligible strangeness in the nucleon. 

\item With both the $\Delta$B$\chi$PT and $\s\Delta$B$\chi$PT approaches we obtain consistent
descriptions of the
  pion mass dependence of the nucleon mass, as can be seen in Figs. \ref{fig:sigma piN
nf21} and \ref{fig:nf2 and nf21 comp}. Moreover, for the current lQCD data, all our
results are compatible within uncertainties and exhibit only small slope variations.
However, these small variations translate into differences in the
  value of $\sigma_{\pi N}$ at the physical point.  For the $2$ and $2+1$ flavor
ensembles the $M_\pi$ distribution of the data points is different.
To further reduce the uncertainty in the
$\sigma_{\pi N}$ value, lQCD data points with smaller  uncertainties and less spread
would be required. In the $N_{f}=2+1$ case a considerable improvement could be achieved
with a direct measurement of $\sigma_{\pi N}$ for $M_\pi <300$ MeV.
It will be interesting to see how the $N_{f}=2$ and $N_{f}=2+1$ values for $\sigma_{\pi
N}$ will change when both data sets become more homogeneous.

\end{itemize}

\begin{acknowledgments}
  We thank Doug Toussaint for making the new MILC data available to us. We are also grateful to Jos\'e M. Alarc\'on,  Michele Della Morte, Jambul Gegelia, Juan Nieves and Vladimir Pascalutsa for useful discussions. The work by LAR, TL and MVV has been supported by the Spanish Ministerio de Economía y Competitividad and European FEDER funds under Contracts FIS2011-28853-C02-01 and FIS2011-28853-C02-02, Generalitat Valenciana under contract PROMETEO/2009/0090 and the EU Hadron-Physics3 project, Grant No. 283286. JMC acknowledges support from the Science Technology and 
Facilities Council (STFC) under grant ST/J000477/1, the Spanish Government and FEDER
funds under contract FIS2011-28853-C02-01 and the grants FPA2010-17806 and Fundaci\'on
S\'eneca 11871/PI/09.

\end{acknowledgments}
\begin{appendix}

\section{B$\chi$PT Lagrangians\label{Appendix: BChPT-Lagrangians}}

The counting scheme of Eq. (\ref{eq: Power Counting}) defines the
nucleon $p^{4}$ self-energy by the sum of the graphs shown in Figs.
\ref{fig: p4 nucleon diagrams}. The relevant $SU\left(2\right)$
covariant B$\chi$PT Lagrangians with explicit $\Delta$-isobar degrees
of freedom are: 
\begin{eqnarray}
  \mathcal{L}_{N} & = & \mathcal{L}_{N\pi}^{\left(1\right)}+\mathcal{L}_{N\Delta\pi}^{\left(1\right)}+\mathcal{L}_{\pi}^{\left(2\right)}+\mathcal{L}_{N\pi}^{\left(2\right)}+\mathcal{L}_{\Delta}^{\left(2\right)}+\mathcal{L}_{N\pi}^{\left(4\right)}\,\,\,\,,\\
  \mathcal{L}_{\Delta} & = & \mathcal{L}_{\Delta\pi}^{\left(1\right)}+\mathcal{L}_{N\Delta\pi}^{\left(1\right)}+\mathcal{L}_{\pi}^{\left(2\right)}\,\,\,\,,
\end{eqnarray}
where the upper indices denote the chiral order. Explicitly, the individual
isospin symmetric Lagrangians in absence of external fields and expanded
in pion fields $\pi$ are: 
\begin{eqnarray}
  \mathcal{L}_{N\pi}^{\left(1\right)} & = & \overline{N}\left[i\s\partial-M_{0}+\frac{1}{4f_{\pi0}^{2}}\epsilon^{abc}\left(\s\partial\pi^{a}\right)\pi^{b}\tau^{c}-\frac{g_{A0}}{2f_{\pi0}}\gamma^{\mu}\gamma^{5}\left(\partial_{\mu}\pi^{a}\right)\tau^{a}\right]N\,\,\,\,,\\
  \mathcal{L}_{\Delta\pi}^{\left(1\right)} & = & \overline{\Delta}_{\mu}\left(\gamma^{\mu\nu\alpha}i\partial_{\alpha}-M_{\Delta0}\gamma^{\mu\nu}\right)\Delta_{\nu}+\frac{H_{A}}{2f_{\pi0}M_{\Delta0}}\varepsilon^{\mu\nu\alpha\lambda}\overline{\Delta}_{\mu}\mathcal{T}^{a}\left(\partial_{\alpha}\Delta_{\nu}\right)\partial_{\lambda}\pi^{a}\,\,\,\,,\\
  \mathcal{L}_{\Delta N\pi}^{\left(1\right)} & = & i\frac{h_{A}}{2f_{\pi0}M_{\Delta0}}\overline{N}T^{a}\gamma^{\mu\nu\lambda}\left(\partial_{\mu}\Delta_{\nu}\right)\partial_{\lambda}\pi^{a}+\mbox{H.c.}\,\,\,\,,\\
  \mathcal{L}_{\pi}^{\left(2\right)} & = & \frac{1}{2}\left(\partial_{\mu}\pi^{a}\right)\left(\partial^{\mu}\pi^{a}\right)-\frac{1}{2}M^{2} \pi^a \pi^a\,\,\,\,,\\
  \mathcal{L}_{N\pi}^{\left(2\right)} & = & c_{1}2m_{\pi}^{2}\left[2-\frac{1}{f_{\pi0}^{2}} \pi^a \pi^a\right]\overline{N}N-\frac{c_{2}}{M_{0}^{2}f_{\pi0}^{2}}\overline{N}\left(\partial_{\mu}\pi^{a}\right)\left(\partial_{\nu}\pi^{a}\right)\partial^{\mu}\partial^{\nu}N\,\,\,\,,\\
  &  & +\frac{c_{3}}{f_{\pi0}^{2}}\left(\partial_{\mu}\pi^{a}\right)\left(\partial^{\mu}\pi^{a}\right)\overline{N}N-\frac{c_{4}}{4f_{\pi0}^{2}}\overline{N}\gamma^{\mu}\gamma^{\nu}\left[\partial_{\mu} \pi^a,\partial_{\nu} \pi^a\right]N+c_{5}\frac{m_{\pi}^{2}}{f_{\pi0}^{2}}\overline{N}\left[\pi^a \pi^a - \left(\pi^a \tau^a\right)^{2}\right]N\,\,\,\,,\nonumber \\
  \mathcal{L}_{\Delta}^{\left(2\right)} & = & 4c_{1\Delta}m_{\pi}^{2}\overline{\Delta}_{\mu}\gamma^{\mu\nu}\Delta_{\nu}\,\,\,\,,\\
  \mathcal{L}_{N\pi}^{\left(4\right)} & = & -\frac{1}{2}\alpha m_{\pi}^{4}\overline{N}N\,\,\,\,,
\end{eqnarray}
where $m_{\pi}^{2}$ is the $\mathcal{O}\left(p^{2}\right)$ pion
mass $m_{\pi}^{2}=2B\overline{m}$ proportional to the chiral condensate
$B$ and the current-quark mass average $\overline{m}$. The Lagrangians
$\mathcal{L}_{N\pi}^{\left(1,2,4\right)}$ for the nucleon field $N$
are those of \cite{Fettes:2000gb} with $\alpha=-4\left[8e_{38}+e_{115}+e_{116}\right]$
a combination of $\mathcal{L}_{N\pi}^{\left(4\right)}$ low energy
constants; the $\mathcal{L}_{N\pi}^{\left(3\right)}$ does not produce
any nucleon self-energy vertices. The couplings of the $\Delta$-isobar
are chosen to be consistent with the covariant construct of the free
Rarita-Schwinger theory and hence do not contain the unphysical degrees
of freedom of vector-spinor fields. The $\Delta$-isobar Lagrangians
and further details can be found in \cite{Pascalutsa:2005nd,Pascalutsa:2006up,Pascalutsa:1998pw,Pascalutsa:1999zz,Pascalutsa:2000kd}.
There are 13 low energy constants $f_{\pi0},g_{A0},c_{1},c_{2},c_{3},c_{4},c_{5},H_{A0},h_{A0},M_{0},M_{\Delta0},c_{1\Delta},\alpha$
where $c_{4}$ and $c_{5}$ do not contribute to the nucleon mass.

The loop graphs in Fig. \ref{fig: p4 nucleon diagrams} are divergent
in $4$ dimensions and need to be regularized. For that we use the
dimensional regularization with $D=4-2\epsilon$ dimensions and renormalize
contributions proportional to: 

\[
L=-\frac{1}{\varepsilon}+\gamma_{E}-\ln4\pi\,\,\,\,.
\]

For the $D$-dimensional spin-$3/2$ propagator we use: 
\[
S_{\Delta}^{\alpha\beta}(p)=\frac{\s p+M_{\Delta}}{p^{2}-M_{\Delta}^{2}+i\varepsilon}\left[-g^{\alpha\beta}+\frac{1}{D-1}\gamma^{\alpha}\gamma^{\beta}+\frac{1}{\left(D-1\right)M_{\Delta}}(\gamma^{\alpha}p^{\beta}-\gamma^{\beta}p^{\alpha})+\frac{D-2}{\left(D-1\right)M_{\Delta}^{2}}p^{\alpha}p^{\beta}\right]\,\,\,.
\]
The appearing totally anti-symmetric $\gamma$-matrices are:

\begin{eqnarray*}
  \gamma^{\mu\nu} & = & \frac{1}{2}\left[\gamma^{\mu},\gamma^{\nu}\right]\,\,\,,\\
  \gamma^{\mu\nu\rho} & = & \frac{1}{2}\left\{ \gamma^{\mu\nu},\gamma^{\rho}\right\} =i\varepsilon^{\mu\nu\rho\sigma}\gamma_{5}\gamma_{\sigma}=\gamma^{\mu\nu\rho\sigma}\gamma_{\sigma}\,\,\,,\\
  \gamma^{\mu\nu\rho\sigma} & = & \frac{1}{2}\left[\gamma^{\mu\nu\rho},\gamma^{\sigma}\right]=i\varepsilon^{\mu\nu\rho\sigma}\gamma_{5}\,\,\,.
\end{eqnarray*}

\section{Self-energy formulas \label{Appendix: Self-energy-formulas}}

\subsection{Nucleon self-energies}

For the nucleon mass we need the self-energy expressions corresponding
to the Feynman-graphs in Fig. \ref{fig: p4 nucleon diagrams}. The
contributions listed in increasing chiral order are:

\begin{eqnarray*}
  \Sigma^{\left(2\right)}\left(m_{\pi}^{2}\right) & = & \Sigma_{C2}\left(m_{\pi}^{2}\right)\,\,\,\,,\\
  \Sigma^{\left(3\right)}\left(m_{\pi}^{2},\s p\right) & = & \Sigma_{N3}\left(m_{\pi}^{2},\s p\right)+\Sigma_{N\Delta3}\left(m_{\pi}^{2},\s p\right)\,\,\,\,,\\
  \Sigma^{\left(4\right)}\left(m_{\pi}^{2},\s p\right) & = & \Sigma_{N4}\left(m_{\pi}^{2},\s p\right)+\Sigma_{T4}\left(m_{\pi}^{2}\right)+\Sigma_{C4}\left(m_{\pi}^{2}\right)+\Sigma_{N\Delta4}\left(m_{\pi}^{2},\s p\right)\,\,\,\,,
\end{eqnarray*}
where we keep the $\s p$ dependence explicit and a '$\Delta$' in
the index denotes contributions from loop-internal $\Delta$-isobars.
The individual unregularized self-energies read:

\begin{eqnarray}
  \Sigma_{C2}\left(m_{\pi}^{2}\right) & = & -c_{1}4m_{\pi}^{2}\\
  \Sigma_{N3}\left(m_{\pi}^{2},\s p\right) & = & 3\left[\frac{g_{A0}}{8f_{\pi0}\pi}\right]^{2}\int_{0}^{1}dz\Big\{\left(z\s p-M_{0}-2\s p\right)\mathcal{M}_{N}^{2}+\left(1-z\right)^{2}\left(\s p\right)^{2}\left(z\s p-M_{0}\right)\left[L+\ln\frac{\mathcal{M}_{N}^{2}}{\Lambda^{2}}\right]\nonumber \\
  &  & +\left(-4\s p-2M_{0}+3z\s p\right)\mathcal{M}_{N}^{2}\left[L-1+\ln\frac{\mathcal{M}_{N}^{2}}{\Lambda^{2}}\right]\Big\}\\
  \Sigma_{N4}\left(m_{\pi}^{2}\right) & = & -c_{1}4m_{\pi}^{2}\left.\frac{\partial}{\partial M_{0}}\Sigma_{N3}\left(m_{\pi}^{2},\s p\right)\right|_{\s p=M_{0}}\\
  & = & c_{1}m_{\pi}^{2}12\left[\frac{g_{A}}{8F_{\pi}\pi}\right]^{2}\int_{0}^{1}dz2\left(1-z\right)\Big\{3\mathcal{M}_{N}^{2}\left[L-1+\ln\frac{\mathcal{M}_{N}^{2}}{\Lambda^{2}}\right]\\
  &  & +3M_{0}^{2}\left(2-2z+z^{2}\right)\left[L+\ln\frac{\mathcal{M}_{N}^{2}}{\Lambda^{2}}\right]+\left(\left(1-z\right)^{2}+2\right)M_{0}^{2}+\frac{5}{2}\mathcal{M}_{N}^{2}+\frac{\left(1-z\right)^{4}}{2\mathcal{M}_{N}^{2}}\Big\}\nonumber \\
  \Sigma_{T4}\left(m_{\pi}^{2}\right) & = & \frac{3}{4F_{\pi}^{2}\left(4\pi\right)^{2}}\left(8c_{1}-c_{2}-4c_{3}\right)\left[L-1+\ln\frac{m_{\pi}^{2}}{\Lambda^{2}}\right]m_{\pi}^{4}+c_{2}\frac{3}{8f_{\pi0}^{2}\left(4\pi\right)^{2}}m_{\pi}^{4}\\
  \Sigma_{C4}\left(m_{\pi}^{2}\right) & = & \frac{1}{2}\alpha m_{\pi}^{4}\\
  \Sigma_{N\Delta3}\left(m_{\pi}^{2},\s p\right) & = & \left[\frac{h_{A0}}{8f_{\pi0}M_{\Delta0}\pi}\right]^{2}\int_{0}^{1}dz\left(z\s p+M_{\Delta0}\right)p^{2}\left\{ -2\mathcal{M}_{\Delta}^{2}-2\mathcal{M}_{\Delta}^{2}\left[L-1+\ln\frac{\mathcal{M}_{\Delta}^{2}}{\Lambda^{2}}\right]\right\} \\
  \Sigma_{N\Delta4}\left(m_{\pi}^{2}\right) & = & c_{1\Delta}8m_{\pi}^{2}\left[\frac{h_{A0}}{8f_{\pi0}\pi M_{\Delta0}}\right]^{2}\int_{0}^{1}dz\,\left(1-z\right)M_{0}^{2}\Big\{3\mathcal{M}^{2}\left[L-1+\ln \frac{\mathcal{M}_{N}^{2}}{\Lambda^2}\right]+4\mathcal{M}^{2}\\
  &  & +\left(M_{\Delta0}^{2}+2M_{0}M_{\Delta0}z+M_{0}^{2}z^{2}\right)\left[L+\ln\tilde{\mathcal{M}}^{2}\right]+M_{\Delta0}^{2}+2M_{0}M_{\Delta0}z+M_{0}^{2}z^{2}\Big\}\,\,\,\,,\nonumber 
\end{eqnarray}
with the expression

\begin{eqnarray}
\mathcal{M}_{N}^{2} & = & zm_{\pi}^{2}-z\left(1-z\right)p^{2}+\left(1-z\right)M_{0}^{2}\,\,\,\,,\\
\mathcal{M}_{\Delta}^{2} & = & zm_{\pi}^{2}-z\left(1-z\right)p^{2}+\left(1-z\right)M_{\Delta0}^{2}\,\,\,\,.
\end{eqnarray}

\subsection{$\Delta\left(1232\right)$ self-energies}

In Sec. \ref{sub:pion mass, fit formula} we use the pion mass dependence
of the $\Delta$-isobar to constrain the LEC $c_{1\Delta}$. The $\Delta$-isobar
mass to order $p^{3}$ is 

\begin{equation}
M_{\Delta}^{\left(3\right)}\left(m_{\pi}^{2}\right)=M_{\Delta0}+\Sigma_{\Delta2}\left(m_{\pi}^{2}\right)+\Sigma_{\Delta N3}\left(m_{\pi}^{2}\right)+\Sigma_{\Delta\Delta3}\left(m_{\pi}^{2}\right)\,\,\,\,,\label{eq: Delta p3 mass}
\end{equation}
where the self-energies are defined as 
\begin{equation}
\Sigma_{\Delta}^{\alpha\beta}\left(\s p\right)=-g^{\alpha\beta}\left[\s p\Sigma_{\Delta}^{A}\left(M_{\Delta0}\right)+\Sigma_{\Delta}^{B}\left(M_{\Delta0}\right)\right]\,\,\,\,,
\end{equation}
with the unregularized expressions

\begin{eqnarray}
\Sigma_{C\Delta2}\left(m_{\pi}^{2}\right) & = & -c_{1\Delta}4m_{\pi}^{2}\,\,\,\,,\\
\Sigma_{\Delta N3}\left(m_{\pi}^{2}\right) & = & -\frac{1}{2}\left[\frac{h_{A}}{8f_{\pi0}\pi}\right]^{2}\int_{0}^{1}dz\left\{ \left(zM_{\Delta0}+M_{0}\right)\mathcal{M}_{\Delta N}^{2}\left[L-1+\ln\frac{\mathcal{M}_{\Delta N}^{2}}{\Lambda^{2}}\right]+4\left(zM_{\Delta0}+M_{N0}\right)\mathcal{M}_{\Delta N}^{2}\right\} \,\,\,\,,\\
 &  & \mathcal{M}_{\Delta N}^{2}=zm_{\pi}^{2}-z\left(1-z\right)M_{\Delta0}^{2}+\left(1-z\right)M_{0}^{2}\,\,\,\,,\\
\Sigma_{\Delta\Delta3}\left(m_{\pi}^{2}\right) & = & -\frac{5}{3}\left[\frac{H_{A}}{8f_{\pi0}\pi}\right]^{2}\int_{0}^{1}dz\left\{ \frac{5}{6}M_{\Delta0}\left(1+z\right)\mathcal{M}_{\Delta\Delta}^{2}\left[L-1+\ln\frac{\mathcal{M}_{\Delta\Delta}^{2}}{\Lambda^{2}}\right]+\frac{13}{9}M_{\Delta0}\left(1+z\right)\mathcal{M}_{\Delta\Delta}^{2}\right\} \,\,\,\,,\\
 &  & \mathcal{M}_{\Delta\Delta}^{2}=zm_{\pi}^{2}-z\left(1-z\right)M_{\Delta0}^{2}+\left(1-z\right)M_{\Delta0}^{2}\,\,\,\,.
\end{eqnarray}
These contributions are the $\Delta$-isobar versions of the nucleon
graphs $\Sigma_{C2}$, $\Sigma_{N3}$ and $\Sigma_{N\Delta3}$ of
Fig. \ref{fig: p4 nucleon diagrams}. The $\Sigma_{C\Delta2}$ is
the $\Delta$-isobar contact graph and the $\Sigma_{\Delta N3}$ and
$\Sigma_{\Delta\Delta3}$ are $p^{3}$ loop with external $\Delta$-isobars
and an internal nucleon and $\Delta$-isobar, respectively. 

\subsection{Finite volume corrections to the nucleon self-energies\label{Appendix: Finite-volume-expressions}}

The loop graphs $\Sigma_{N3}$, $\Sigma_{N4}$, $\Sigma_{T4}$ and
$\Sigma_{N\Delta3}$, $\Sigma_{N\Delta4}$ of Fig. \ref{fig: p4 nucleon diagrams}
are subject to finite volume (FV) effects when the nucleon is placed
in a discretized box. We calculate these effects by the standard techniques
of \cite{AliKhan:2003cu}. In the following we summarize the calculation
of the loop-integral with a single propagator and list afterwards
all appearing FV corrections for the nucleon mass to order $p^{4}$.

For the FV calculation we chose the nucleon rest-frame $\s p=\gamma_{0}p_{0}=\gamma_{0}M_{N}$.
As a consequence all appearing loop-integrals can be brought into
the form of 
\begin{equation}
\int\frac{dl^{4}}{\left(2\pi\right)^{4}}\frac{l.A\, l.B\,\cdots}{l^{2}-m^{2}}\to\int\frac{dl^{4}}{\left(2\pi\right)^{4}}\frac{l_{0}^{a}}{l^{2}-m^{2}}\,\,\,\,,
\end{equation}
where no Lorentz-decomposition has to be used, $A$ and $B$ are given
4-vectors and $a$ a power of the 0th-loop momentum component. The
loop-momentum $l$ is now discretized with respect to the box size
$L$ by

\begin{eqnarray}
\int\frac{d^{4}l}{\left(4\pi\right)^{4}}=\int\frac{dl_{0}}{2\pi}\frac{d\vec{l}}{\left(2\pi\right)^{3}} & \to & \int\frac{dl_{0}}{2\pi}\frac{1}{L^{3}}\sum_{\vec{n}}\,\,\,\,\mbox{with}\,\,\,\vec{l}=\frac{2\pi}{L}\vec{n}\,\,\,\,\vec{n}\in\mathbb{Z}^{3}\,\,\,,
\end{eqnarray}
such that after Wick-rotating and the use of Poisson's formula we
get:

\begin{eqnarray}
\int\frac{dl_{0}}{2\pi}\frac{1}{L^{3}}\sum_{\vec{n}}\frac{l_{0}^{a}}{l_{0}^{2}-\frac{2\pi}{L}\vec{n}^{2}-m^{2}} & = & -i^{\alpha+1}\int_{-\infty}^{\infty}\frac{dl_{4}}{2\pi}\int_{-\infty}^{\infty}\frac{d\vec{l}}{\left(2\pi\right)^{3}}\frac{l_{4}^{a}}{l_{4}^{2}+\vec{l}^{2}+m^{2}}\frac{\left(2\pi\right)^{3}}{L^{3}}\sum_{\vec{n}}\delta^{\left(3\right)}\left(\vec{l}-\frac{2\pi}{L}\vec{n}^{2}\right)\\
 & = & -i^{\alpha+1}\int_{-\infty}^{\infty}\frac{dl_{4}}{2\pi}\int_{-\infty}^{\infty}\frac{d\vec{l}}{\left(2\pi\right)^{3}}\frac{l_{4}^{a}}{l_{4}^{2}+\vec{l}^{2}+m^{2}}\sum_{\vec{j}}e^{iL\vec{j}\cdot\vec{l}}\,\,\,\,,
\end{eqnarray}
with $\vec{j}\in\mathbb{Z}^{3}$. The case $\vec{j}=0$ corresponds to the usual continuum result whereas
the cases $\vec{j}\neq0$ are the finite volume corrections. All remaining
integrals can be solved analytically. For our nucleon mass expression
we need the following solutions:

\begin{eqnarray}
\int\frac{dl^{4}}{\left(2\pi\right)^{4}}\frac{1}{l^{2}-m^{2}}=\frac{-i}{\left(4\pi\right)^{2}}\sum_{\vec{j}\neq0}4\frac{\sqrt{m^{2}}}{Lj}K_{1}\left(F\right) & \,\,\,\,,\,\,\,\, & \int\frac{dl^{4}}{\left(2\pi\right)^{4}}\frac{l_{0}^{2}}{l^{2}-m^{2}}=\frac{-i}{\left(4\pi\right)^{2}}\sum_{\vec{j}\neq0}\frac{\left(-4\right)m^{2}}{\left(Lj\right)^{2}}K_{2}\left(F\right)\,\,\,\,,\\
\int\frac{dl^{4}}{\left(2\pi\right)^{4}}\frac{1}{\left[l^{2}-m^{2}\right]^{2}}=\frac{-i}{\left(4\pi\right)^{2}}\sum_{\vec{j}\neq0}\left(-2\right)K_{0}\left(F\right) & \,\,\,\,,\,\,\,\, & \int\frac{dl^{4}}{\left(2\pi\right)^{4}}\frac{l_{0}^{2}}{\left[l^{2}-m^{2}\right]^{2}}=\frac{-i}{\left(4\pi\right)^{2}}\sum_{\vec{j}\neq0}2\frac{\sqrt{m^{2}}}{Lj}K_{1}\left(F\right)\,\,\,\,,\\
\int\frac{dl^{4}}{\left(2\pi\right)^{4}}\frac{1}{\left[l^{2}-m^{2}\right]^{3}}=\frac{-i}{\left(4\pi\right)^{2}}\sum_{\vec{j}\neq0}\frac{1}{2}\frac{Lj}{\sqrt{m^{2}}}K_{1}\left(F\right) & \,\,\,\,,\,\,\,\, & \int\frac{dl^{4}}{\left(2\pi\right)^{4}}\frac{l_{0}^{2}}{\left[l^{2}-m^{2}\right]^{3}}=\frac{-i}{\left(4\pi\right)^{2}}\sum_{\vec{j}\neq0}\left(-\frac{1}{2}\right)K_{0}\left(F\right)\,\,\,\,,
\end{eqnarray}
where the $K_{\nu}\left(x\right)$ are modified Bessel-functions of
the second kind with $F=Lj\sqrt{m^{2}}$ and $j=\sqrt{j_{x}+j_{y}+j_{z}}$
with $j_{i}\in Z$. 

To collect our final results we use the notations:
\begin{eqnarray}
F_{N}=Lj\sqrt{\mathcal{M}_{N}^{2}} & \,\,\,\,,\,\,\,\, & \Sigma_{N3}^{\prime}\left(m_{\pi}^{2},L\right)=\left.\frac{\partial}{\partial p_{0}}\Sigma_{N3}\left(p_{0},m_{\pi}^{2},L\right)\right|_{p_{0}=M_{0}}\,\,\,\,,\\
F_{\Delta}=Lj\sqrt{\mathcal{M}_{\Delta}^{2}} & \,\,\,\,,\,\,\,\, & \Sigma_{N\Delta3}^{\prime}\left(m_{\pi}^{2},L\right)=\left.\frac{\partial}{\partial p_{0}}\Sigma_{N\Delta3}\left(p_{0},m_{\pi}^{2},L\right)\right|_{p_{0}=M_{0}}\,\,\,\,,
\end{eqnarray}
where the arguments of the self-energies distinguish them from their
continuum counterparts.

The individual finite volume contributions corresponding to the loop-graphs
in Fig. \ref{fig: p4 nucleon diagrams} are:

\begin{eqnarray}
\Sigma_{N3}\left(m_{\pi}^{2},L\right) & = & 3\left[\frac{g_{A}}{8f_{\pi}\pi}\right]^{2}\sum_{\vec{j}\neq0}\int_{0}^{1}dz2M_{0}\Big[\nonumber  \\
 &  & \left(\left(1-z\right)^{3}M_{0}^{2}+\left(3-z\right)\mathcal{M}^{2}\right)K_{0}\left(F_{N}\right)+\left(4z-6\right)\frac{\sqrt{\mathcal{M}_N^{2}}}{Lj}K_{1}\left(F_{N}\right)\Big] \\
\Sigma_{N3}^{\prime}\left(m_{\pi}^{2},L\right) & = & 3\left[\frac{g_{A}}{8f_{\pi}\pi}\right]^{2}\sum_{\vec{j}\neq0}\int_{0}^{1}dz\,\,\,\,4\Big[+\left(2z-2\right)\frac{\sqrt{\mathcal{M}_N^{2}}}{Lj}K_{1}\left(F_{N}\right)\nonumber \\
 &  & -\frac{1}{2}\left(\left(z-2\right)\mathcal{M}_N^{2}+M_{0}^{2}\left(1-z\right)\left[\left(1-z\right)\left(3z-2\right)-2z^{2}-4z\left(z-3\right)\right]\right)K_{0}\left(F_{N}\right)\nonumber \\
 &  & +\frac{1}{2}z\left(1-z\right)M_{0}^{2}\left(\left(1-z\right)^{3}M_{0}^{2}+\left(3-z\right)\mathcal{M}_N^{2}\right)\frac{Lj}{\sqrt{\mathcal{M}_{N}^{2}}}K_{1}\left(F_{N}\right)\,\,\,\,\Big]\\
\Sigma_{N4}\left(m_{\pi}^{2},L\right) & = & -c_{1}4m_{\pi}^{2}\,\,\,3\left[\frac{g_{A}}{8f_{\pi}\pi}\right]^{2}\int_{0}^{1}dz\sum_{\vec{j}\neq0}\,\,\,\,\Big[\\
 &  & +2\left(\mathcal{M}_{N}^{2}+\left(1-z\right)^{2}M_{0}^{2}-2z\left(1-z\right)M_{0}^{2}-4\left(1-z\right)\left(z-3\right)M_{0}^{2}\right)K_{0}\left(F_{N}\right)\nonumber \\
 &  & +2M_{0}^{2}\left(1-z\right)\left(\left(z-3\right)\mathcal{M}_{N}^{2}-\left(1-z\right)^{3}M_{0}^{2}\right)\frac{Lj}{\sqrt{\mathcal{M}_{N}^{2}}}K_{1}\left(F_{N}\right)-4\frac{\sqrt{\mathcal{M}_{N}^{2}}}{Lj}K_{1}\left(F_{N}\right)\,\,\,\,\Big]\nonumber \\
\Sigma_{T4}\left(m_{\pi}^{2},L\right) & = & \frac{12m_{\pi}^{2}}{F_{\pi}^{2}\left(4\pi\right)^{2}}\sum_{\vec{j}\neq0}\\
 &  & \left[2c_{1}\frac{\sqrt{m_{\pi}^{2}}}{Lj}K_{1}\left(Lj\sqrt{m_{\pi}^{2}}\right)+c_{2}\frac{1}{\left(Lj\right)^{2}}K_{2}\left(Lj\sqrt{m_{\pi}^{2}}\right)-c_{3}\frac{\sqrt{m_{\pi}^{2}}}{Lj}K_{1}\left(Lj\sqrt{m_{\pi}^{2}}\right)\right]\nonumber \\
\Sigma_{N\Delta3}\left(m_{\pi}^{2},L\right) & = & \frac{4}{3}\left[\frac{h_{A}}{8f_{\pi}\pi M_{\Delta0}}\right]^{2}\int_{0}^{1}dz\,\,\left(zM_{0}+M_{\Delta0}\right)2M_{0}^{2}\\
 &  & \left[-\frac{\sqrt{\mathcal{M}_{\Delta}^{2}}}{Lj}K_{1}\left(Lj\sqrt{\mathcal{M}_{\Delta}^{2}}\right)+\mathcal{M}_{\Delta}^{2}K_{0}\left(Lj\sqrt{\mathcal{M}_{\Delta}^{2}}\right)\right]\nonumber \\
\Sigma_{N\Delta3}^{\prime}\left(m_{\pi}^{2},L\right) & = & \frac{4}{3}\left[\frac{h_{A}}{8f_{\pi}M_{\Delta0}\pi}\right]^{2}\int_{0}^{1}dz\Big[\,\,\,\,2z\left(1-z\right)M_{0}^{3}\left(zM_{0}+M_{\Delta0}\right)\mathcal{M}_{\Delta}^{2}\frac{Lj}{\sqrt{\mathcal{M}_{\Delta}^{2}}}K_{1}\left(F_{\Delta}\right)\nonumber \\
 &  & -2M_{0}\left(3zM_{0}+2M_{\Delta0}\right)\frac{\sqrt{\mathcal{M}_{\Delta}^{2}}}{Lj}K_{1}\left(F_{\Delta}\right)\\
 &  & +\left(-6z\left(1-z\right)M_{0}^{3}\left(zM_{0}+M_{\Delta0}\right)+2M_{0}\left(3zM_{0}+2M_{\Delta0}\right)\mathcal{M}_{\Delta}^{2}\right)K_{0}\left(F_{\Delta}\right)\,\,\,\,\Big]\nonumber \\
\Sigma_{N\Delta4}\left(m_{\pi}^{2},L\right) & = & c_{1\Delta}4m_{\pi}^{2}2\left[\frac{h_{A}}{8f_{\pi}\pi M_{\Delta0}}\right]^{2}\int_{0}^{1}dz\,2\left(1-z\right)\frac{1}{3}M_{0}^{2}\,\,\,\,\Big[\\
 &  & -\left(3z^{2}M_{0}^{2}+3M_{\Delta0}^{2}+6zM_{0}M_{\Delta0}+7\mathcal{M}_{\Delta}^{2}\right)K_{0}\left(F_{\Delta}\right)\nonumber \\
 &  & +4\frac{\sqrt{\mathcal{M}_{\Delta}^{2}}}{Lj}K_{1}\left(F_{\Delta}\right)+\mathcal{M}_{\Delta}^{2}\left(z^{2}M_{0}^{2}+M_{\Delta0}^{2}+2zM_{0}M_{\Delta0}+\mathcal{M}_{\Delta}^{2}\right)\frac{Lj}{\sqrt{\mathcal{M}_{\Delta}^{2}}}K_{1}\left(F_{\Delta}\right)\,\,\,\,\Big]\,\,\,\,.\nonumber 
\end{eqnarray}

\subsection{Fit formulas\label{Appendix sub:Fit-formulas}}

In Secs. \ref{sub:Nf2 fits} and \ref{sub:Nf21 fits} we use in the
$\chi^{2}$ fits the following nucleon mass expressions:

\begin{eqnarray}
M_{N}^{\left(2\right)}\left(M_{\pi}^{2}\right) & = & M_{0}+\Sigma_{C2}\left(M_{\pi}^{2}\right)\,\,\,\,,\label{Appendix eq: fit MN2}\\
M_{N}^{\left(3\right)}\left(M_{\pi}^{2}\right) & = & M_{0}+\Sigma_{C2}\left(M_{\pi}^{2}\right)+\Sigma_{N3}\left(M_{\pi}^{2}\right)\,\,\,\,,\label{Appendix eq: fit MN3}\\
M_{N}^{\left(3\Delta\right)}\left(M_{\pi}^{2}\right) & = & M_{0}+\Sigma_{C2}\left(M_{\pi}^{2}\right)+\Sigma_{N3}\left(M_{\pi}^{2}\right)+\Sigma_{N\Delta3}\left(M_{\pi}^{2}\right)\,\,\,\,,\label{Appendix eq: fit MN3D}\\
M_{N}^{\left(4\right)}\left(M_{\pi}^{2}\right) & = & M_{0}+\Sigma_{C2}\left(M_{\pi}^{2}\right)+\Sigma_{N3}\left(M_{\pi}^{2}\right)+\Sigma_{N4}\left(M_{\pi}^{2}\right)+\Sigma_{T4}\left(M_{\pi}^{2}\right)\label{Appendix eq: fit MN4}\\
 &  & +\frac{1}{2}\overline{\alpha}M_{\pi}^{4}+\Sigma_{C2}\left(M_{\pi}^{2}\right)\Sigma_{N3}^{\prime}\left(M_{\pi}^{2}\right)+\frac{c_{1}}{8\pi^{2}f_{\pi}^{2}}M_{\pi}^{4}\ln\frac{M_{\pi}^{2}}{M_{N}^{2}}\,\,\,\,,\nonumber \\
M_{N}^{\left(4\Delta\right)}\left(M_{\pi}^{2}\right) & = & M_{0}+\Sigma_{C2}\left(M_{\pi}^{2}\right)+\Sigma_{N3}\left(M_{\pi}^{2}\right)+\Sigma_{N4}\left(M_{\pi}^{2}\right)+\Sigma_{T4}\left(M_{\pi}^{2}\right)\nonumber \\
 &  & +\frac{1}{2}\overline{\alpha}M_{\pi}^{4}+\Sigma_{C2}\left(M_{\pi}^{2}\right)\Sigma_{N3}^{\prime}\left(M_{\pi}^{2}\right)+\frac{c_{1}}{8\pi^{2}f_{\pi}^{2}}M_{\pi}^{4}\ln\frac{M_{\pi}^{2}}{M_{N}^{2}}\nonumber \\
 &  & +\Sigma_{N\Delta3}\left(M_{\pi}^{2}\right)+\Sigma_{N\Delta4}\left(M_{\pi}^{2}\right)+\Sigma_{C2}\left(M_{\pi}^{2}\right)\Sigma_{N\Delta3}^{\prime}\left(M_{\pi}^{2}\right)\,\,\,\,,\label{Appendix eq: fit MN4D}
\end{eqnarray}
where all loops are evaluated at $\s p=M_{0}$. The additional terms
proportional to $c_{1}$, as compared to Eq. (\ref{eq: Nucleon p4 mass}),
come from the discussion in Sec. \ref{sub:pion mass, fit formula}.
In the case of fits with finite volume corrections, we add the following
expressions:
\begin{eqnarray}
\Sigma_{FV}^{\left(3\right)}\left(M_{\pi}^{2},L\right) & = & \Sigma_{N3}\left(M_{\pi}^{2},L\right)\\
\Sigma_{FV}^{\left(3\Delta\right)}\left(M_{\pi}^{2},L\right) & = & \Sigma_{N3}\left(M_{\pi}^{2},L\right)+\Sigma_{N\Delta3}\left(M_{\pi}^{2},L\right)\\
\Sigma_{FV}^{\left(4\right)}\left(M_{\pi}^{2},L\right) & = & \Sigma_{N3}\left(M_{\pi}^{2},L\right)+\Sigma_{N4}\left(M_{\pi}^{2},L\right)+\Sigma_{T4}\left(M_{\pi}^{2},L\right)+\Sigma_{C2}\left(M_{\pi}^{2}\right)\Sigma_{N3}^{\prime}\left(M_{\pi}^{2},L\right)\\
\Sigma_{FV}^{\left(4\Delta\right)}\left(M_{\pi}^{2},L\right) & = & \Sigma_{N3}\left(M_{\pi}^{2},L\right)+\Sigma_{N\Delta3}\left(M_{\pi}^{2},L\right)\nonumber \\
 &  & +\Sigma_{N4}\left(M_{\pi}^{2},L\right)+\Sigma_{N\Delta4}\left(M_{\pi}^{2},L\right)+\Sigma_{T4}\left(M_{\pi}^{2},L\right)\nonumber \\
 &  & +\Sigma_{C2}\left(M_{\pi}^{2}\right)\Sigma_{N3}^{\prime}\left(M_{\pi}^{2},L\right)+\Sigma_{C2}\left(M_{\pi}^{2}\right)\Sigma_{N\Delta3}^{\prime}\left(M_{\pi}^{2},L\right)
\end{eqnarray}

\end{appendix}


\end{document}